\documentclass{article}
\usepackage{graphicx} 
\usepackage{blindtext}
\usepackage{amsmath}
\usepackage{amsfonts}
\usepackage{graphicx}
\usepackage{tikz-cd}
\usepackage[colorlinks=true, allcolors=blue]{hyperref}

\newcommand{\Z}{{\mathbb{Z}}}

\newcommand{\beq}{\begin{equation}}
\newcommand{\eeq}{\end{equation}}
\newcommand{\beqn}{\begin{eqnarray}}
\newcommand{\eeqn}{\end{eqnarray}}

\newcommand{\cH}{ {\cal H} }

\newcommand{\cL}{ {\cal L} }

\newcommand{\SU}{\mathrm{SU}}
\newcommand{\U}{\mathrm{U}}

\newcommand{\RR}{\mathbb{R}}

\newcommand{\lr}[1]{\lfloor #1 \rceil}

\newcommand{\cz}[1]{{\color{magenta}{[carolyn: #1]}}}
\newcommand{\lf}[1]{{\color{red}{[lukasz: #1]}}}

\newcommand{\Hg}{{\cal{H}}}
\newcommand{\Hmatter}{{\cal{H}}_{\text{matter}}}

\newcommand{\HgN}{{\cal{H}}^N}
\newcommand{\Hge}{{\cal{H}}^e}
\newcommand{\Hgm}{{\cal{H}}^m}
\newcommand{\dd}{\!\ast d\!\ast}

\newcommand{\ve}{\varepsilon}
\newcommand{\vt}{\vartheta}
\usepackage[normalem]{ulem}

\title{Non-invertible bosonic chiral symmetry on the lattice}
\date{October 2025}

\author{Lukasz Fidkowski, Cenke Xu, and Carolyn Zhang}

\begin{document}

\maketitle

\begin{abstract}
In this work we realize the $3+1$d non-invertible $\Z_N$ chiral symmetry generator as an operator in a many body lattice Hilbert space.  A crucial ingredient in our construction is the use of infinite dimensional $U(1)$ rotor site Hilbert spaces.  Specifically, our Hilbert space is that of a $U(1)$ lattice gauge theory coupled to a charge $1$ scalar in the Villain formulation, which allows for direct access to monopoles and for a simple definition of a magnetic $\Z_N$ one-form symmetry $Z_m^{(1)}$, at the lattice Hamiltonian level.  We construct the generator of the $\Z_N$ chiral symmetry as as a unitary operator in the subspace of $Z_m^{(1)}$-invariant states, and show that it cannot be extended to the entire Hilbert space while preserving locality and unitarity.  Using a lattice-level duality based on gauging $Z_m^{(1)}$, we find a dual description of this subspace, as the subspace of a charge $\frac{1}{N}$ gauge theory invariant under an electric one-form symmetry $Z_e^{(1)}$.  We show that in this dual formulation, the chiral symmetry generator does extend unitarily to the entire Hilbert space, but has a mixed anomaly with the $Z_e^{(1)}$ symmetry.

\end{abstract}
\section{Introduction}

Symmetries play an important role in both continuum quantum field theory and in condensed matter many-body quantum lattice systems.  Recently, the notion of symmetry has expanded to include higher-form symmetries, whose defects have co-dimension greater than one \cite{Gaiotto2015}, higher-group symmetries, which mix defects of different co-dimensions \cite{Kapustin2013SPTs, Tachikawa2020, Cordova2019, Benini2019}, and non-invertible symmetries, whose defects do not form a group under fusion \cite{Bhardwaj2018, Chang2019, Kaidi2022, Choi2022NonInvertible,thorngren2024,thorngren20242}.  Among the non-invertible symmetries, one particularly interesting example is that of the discrete non-invertible chiral symmetry in certain $3+1$d $U(1)$ gauge theories \cite{Choi_standard_model, Cordova_Ohmori, Shao_TASI}.  This is a discrete remnant of the $U(1)_A$ axial symmetry that is broken by the Adler-Bell-Jackiw (ABJ) anomaly \cite{Adler1969, BellJackiw1969}.  The corresponding symmetry defect hosts a quantum Hall state with a fractional Hall conductivity.  Such a state must also have anyons with fractional statistics, and it is the non-invertability of this $2+1$d state that, in the field theory interpretation, is the origin of the non-invertible nature of the $3+1$d chiral symmetry \cite{Shao_TASI}.  However, this intuition does not directly lead to an operator-theoretic understanding of the non-invertible chiral symmetry: how does a non-invertible $2+1$d fractional quantum Hall state living on the symmetry defect lead to a non-invertible operator implementing the action of the symmetry in a many body lattice Hilbert space.

In this work we answer this question by explicitly constructing the action of the discrete non-invertible chiral symmetry in a bosonic many body lattice Hilbert space.\footnote{We will work with a branched simplicial decomposition of a spatial manifold $M^3$, but our results will also apply to the ordinary cubic lattice.  The only structure we need is that of higher cup products.}  In our approach we first construct the action of an axial $U(1)_A$ symmetry in the `un-gauged' theory, i.e. one with a global onsite $U(1)_V$ symmetry.  This is non-trivial due to the Nielsen-Ninomiya theorem \cite{NielsenNinomiya1981I, NielsenNinomiya1981II} and its extension to interacting systems \cite{FidkowskiXu}.  However, these no-go theorems can be evaded in various ways, for example by changing the $U(1)_A$ to an $\RR$ symmetry \cite{gioia2025} or by relaxing the requirement that $U(1)_A$ and $U(1)_V$ commute, resulting in an Onsager algebra \cite{ShaoOnsager,pace2025T}.  In this work we will retain $U(1)_A$ as a compact symmetry and insist that it commute with $U(1)_V$, but instead evade the no-go theorems by working with a lattice of infinite dimensional rotor degrees of freedom.\footnote{See also ref. \cite{ChengSeibergLSM} for a similar result in $1+1$d.}  In this setting, the onsite $U(1)_V$ simply shifts the rotors, while the $U(1)_A$ multiplies by a certain non-onsite rotor-dependent phase, given as a sum of a local charge density $*j_A^{\text{matter}}$.  The mixed `t Hooft anomaly that characterizes these symmetries is manifested by the fact that acting with a $2\pi$ $U(1)_A$ rotation pumps a $U(1)$ bosonic symmetry protected topological (SPT) state \cite{SenthilLevin2013}, characterized by non-zero Hall conductivity, out to the boundary.  Indeed, in the case of the standard axial symmetry of a Dirac fermion, a $2\pi$ axial rotation pumps out an integer quantum Hall (IQH) state to the boundary; equivalently, a fundamental vortex in the Dirac mass term hosts a $c=1$ chiral edge mode \cite{FidkowskiXu}, which is the obstruction to obtaining a gapped $U(1)_A$-symmetric by quantum disordering the Dirac mass order parameter.\footnote{Another useful analogy is to a $1+1$ dimensional Dirac fermion, where a $2\pi$ axial rotation acts as a translation that pumps charge (the so-called Thouless pump~\cite{thouless,solitoncharge}), i.e. a $0$ dimensional SPT.} 
Our lattice situation is different in two ways: we have a bosonic system instead of a fermionic one, and our $U(1)_A$ symmetry has no self-anomaly.  The latter property is reflected by the fact that we pump an SPT phase, which has no net chiral central charge, only $U(1)$ Hall conductivity.  We provide a field theory realization of this anomaly in section \ref{subsec:target}.  The ability of a many body Hilbert space made up of infinite dimensional rotor Hilbert spaces on each site to evade the no-go theorems of \cite{NielsenNinomiya1981I, NielsenNinomiya1981II, FidkowskiXu} is closely tied to the existence of $2+1$ d commuting projector Hamiltonians in rotor Hilbert spaces with non-zero Hall conductivity \cite{DeMarco_SPT}, which evade the no-go theorems of \cite{KapustinFidkowski2019, ZhangLevin2022}.

The second step in our approach is to gauge the $U(1)_V$ symmetry by coupling to a $U(1)_V$ gauge field, at the level of the spatial lattice.  It will be important for us to retain control over the monopoles, so we use the Villain formalism where the $U(1)_V$ vector potential $a$ is real-valued and defined on the edges of our spatial lattice, and the integer-valued magnetic variable $n$ is defined on the plaquettes; the real-valued rotors $\phi$ live on the vertices.\footnote{The compactness of $\phi$, i.e. the invariance of the wavefunction under integer shifts of $\phi$, follows from the Villain condition and gauge invariance.}  The Hilbert space $\Hg$ consists of all wavefunctions of $\phi,a,n$ that are invariant under gauge invariance (eq. \ref{usual_gauge_invariance}) and the Villain condition (eq. \ref{Villain}).  The latter can be thought of as a gauge invariance condition arising from gauging a $\mathbb{Z}$ subgroup of the $\mathbb{R}$ one-form symmetry implementing shifts of $a$, with $n$ acting as the $\mathbb{Z}$ two-form gauge field.  The matter rotor Hilbert space can be viewed as a subspace $\Hmatter \subset \Hg$ where we have made the gauge field trivial, i.e. set $a=d\lambda + m$, $n=dm$, where $\lambda$ is real and $m$ is integer-valued.  We then write down a local charge density $*j_A$ which may be thought of as a `gauging' of $*j_A^{\text{matter}}$, in that it agrees with $*j_A^{\text{matter}}$ on $\Hmatter$.  Our $*j_A$ will not generate a $U(1)$ symmetry, but it will have the property that
\begin{align*}
\exp & \left[2\pi i \int_{M^3} *j_A\right] = \nonumber \\ &\exp \left[2\pi i\int_{M^3}\left(da \cup a - a\cup n - n\cup a - a \cup_1 dn - 2\phi\cup dn\right)\right],
\end{align*}
where $M^3$ is the spatial manifold, and the integral is really a sum over $3$-simplices.  The lattice-level cup products in this equation are defined for both branched simplicial decompositions (see e.g. appendix A of \cite{Ellison2021}) and for ordinary cubic lattices \cite{ChenTata2023}, so we can work with either kind of lattice.  The expression on the right hand side of the above equation is both gauge invariant and invariant under the Villain condition \cite{Jacobson}.  It has previously been used to define a $2+1$d action for lattice Chern-Simons theory \cite{Jacobson, DeMarco_CS}, but here it is interpreted as an operator in a $3+1$d theory.  The above equation shows that $*j_A$ generates a family of local unitaries parametrized by $\RR$, as opposed to $U(1)$, and also that these unitaries fail to be symmetries for any Hamiltonian that includes an electric field operator.  This is the expected anomalous non-conservation of $U(1)_A$ in the presence of a $U(1)_V$ gauge field.

Since we are interested in constructing only a $\Z_N$ subgroup of $U(1)_A$, however, we could attempt to define the generator of this $\Z_N$ group by
\begin{align} \label{tentative_gen}
\exp\left[\frac{2\pi i}{N}\int_{M^3} \left(*j_A-(da \cup a - a\cup n - n\cup a - a \cup_1 dn - 2\phi\cup dn) \right)\right]
\end{align}
The $N$'th power of this operator is the identity by construction, and for a gauge field configuration with no monopoles ($dn=0$), the only $\phi$ dependence in the above exponent is in $*j_A$. Thus this operator is equal to $\exp\left[\frac{2\pi i}{N}\int_{M^3} *j_A\right]$ on $\Hmatter$, i.e. it has the correct axial action on the matter degrees of freedom.  However, the problem is that this is not actually a well defined operator in $\Hg$, because the expression in the exponent is not invariant under the Villain constraint (eq. \ref{Villain}).  Indeed, this can immediately be seen from the fact that the operator in eq. \ref{tentative_gen} is not invariant under integer shifts of $\phi$.

To solve this problem, we will temper our ambition and only attempt to define a $\Z_N$ generator on a certain subspace $\Hgm$ of $\Hg$.  Our $\Z_N$ generator will then involve a projection onto this subspace, which will be the origin of its non-invertible nature.  The subspace $\Hgm$ will be the set of states invariant under the $\Z_N$ magnetic $1$-form symmetry
\begin{align*}
Z_m^{(1)}(\Sigma)\cdot \Psi(\phi,a,n) = \exp\left[\frac{2 \pi i}{N}\int_\Sigma n\right] \Psi(\phi,a,n),
\end{align*}
where $\Sigma$ is a homology $2$-surface in $M^3$:
\begin{align*}
\Hgm = \{|\psi\rangle \in \Hg \,| \,Z_m^{(1)}(\Sigma)\cdot|\psi\rangle = |\psi\rangle \text{ for all } \Sigma\}
\end{align*}
Again, since we are working on a lattice, all integrals are to be interpreted as sums, so that the integral over $\Sigma$ is really a sum over the $2$-simplices making up $\Sigma$.  Invariance under $\Z_m^{(1)}(\Sigma)$ for all $\Sigma$ implies that $n$ is an exact $2$-cocycle when viewed modulo $N$, i.e. that there exists an integer-valued $1$-cochain $\alpha$ such that $n=d\alpha$ mod $N$. We will also use the notation $\alpha = d^{-1}n$.  $d^{-1}n$ is unique up to shifts by integer valued $1$-cochains which are closed modulo $N$.  We then define
\begin{align} \label{defU_intro}
U_{\frac{1}{N}}&(M^3)= \nonumber \\&\exp\left[\frac{2\pi i}{N}\int_{M^3} \left(*j_A-((da-n) \cup (a-d^{-1}n)  - a \cup_1 dn - 2\phi\cup dn) \right)\right] \cdot P_{\Hgm}
\end{align}
where $P_{\Hgm}$ is the projection onto $\Hgm$.  This is now a well-defined operator in $\Hg$ (in particular, note that $dn \in N\Z$ in $\Hgm$, so the operator is invariant under integer shifts of $\phi$), and is our definition of the non-invertible chiral symmetry operator.  Since on configurations with no monopoles ($dn=0$) the only $\phi$ dependence is in $*j_A$, and $\Hmatter$ is in the image of the projector $P_{\Hgm}$, $U_{\frac{1}{N}}(M^3)$ performs the requisite invertible axial rotation on the matter degrees of freedom.  

The expression subtracted from $*j_A$ in eq. \ref{defU_intro}, together with the factor of $\frac{1}{N}$, may be thought of as a particular way of defining a gauge invariant fractional Chern-Simons term ``$\frac{1}{N} a\,da$", on the spatial lattice.  However, the connection between this expression and the topological quantum field theory (TQFT) based method of curing the ABJ anomaly \cite{Shao_TASI} is not immediately clear.  To elucidate this connection, we gauge the $\Z_N$ 1-form symmetry $Z_m^{(1)}$ by minimally coupling to a $2$-form spatial lattice $\Z_N$ gauge field.  Similarly to what happens in simpler examples in $1+1$d \cite{KWAasen, KWSeiberg1, KWSeiberg2} and $3+1$d \cite{threeDApte2023, threeDHsin2019comments, Choi2022NonInvertible, Kaidi2022, gorantla2024tensor, zhang2024anomalies}, such gauging gives a locality-preserving isomorphism between $\Hgm$ and another Hilbert space, $\Hge \subset \HgN$, where $\HgN$ is the Villain Hilbert space of a charge $\frac{1}{N}$ scalar coupled to a $U(1)$ gauge field, and $\Hge$ is the subspace invariant under an electric $1$-form symmetry $Z_e^{(1)}$ which effectively just measures the fractional part of the charge of a Wilson line.  Intuitively, $\Hgm \cong \Hge$ because the operator algebras of the two Hilbert spaces match: in addition to the electric and magnetic fields, both contain integer charge Wilson lines and magnetic charge $N$ `t Hooft lines.

In summary, we consider four Hilbert spaces:
\begin{itemize}
\item $\mathcal{H}$ is a Hilbert space of an integer charge scalar coupled to a $U(1)$ gauge field via the Villain formalism
\item $\mathcal{H}^m\subset \mathcal{H}$ is the subspace of $\mathcal{H}$ symmetric under the magnetic 1-form symmetry, which only allows for magnetic charge $N$ monopoles
\item $\mathcal{H}^N$ is a Hilbert space with a different Villain condition, resulting in a period $N$ scalar carrying electric charge $1/N$ coupled to a $U(1)$ gauge field with magnetic charge $N$ monopoles
\item $\mathcal{H}^e\subset \mathcal{H}^N$ is the subspace of $\mathcal{H}^N$ invariant under the electric 1-form symmetry, which removes the fractional electric charges
\end{itemize}

We can then pull back $U_{\frac{1}{N}}(M^3)$ defined in eq. \ref{defU_intro} to $\Hge$.  Using the explicit form of $*j_A$ we find that 
\begin{align} \label{U_in_Hge}
U_{\frac{1}{N}}(M^3) = \exp\left[-\frac{2\pi i}{N}\int_{M^3} \lr{a-d\phi}\cup d\lr{a-d\phi}\right] \cdot P_{\Hge}
\end{align}
where by abuse of notation we also denote this pullback by $U_{\frac{1}{N}}(M^3)$.  Here $\lr{x}$ denotes the integer closest to the real number $x$.  As opposed to the action of $U_{\frac{1}{N}}(M^3)$ in $\Hgm$, which cannot be extended in a locality-preserving way to all of $\Hg$, its action in $\Hge$, as defined in eq. \ref{U_in_Hge}, can be extended to all of $\HgN$ - in other words, $U_{\frac{1}{N}}(M^3)$ can be written as an invertible operator in $\mathcal{H}^N$ but is necessarily non-invertible in $\mathcal{H}$.  In fact, this extension to $\mathcal{H}^N$, given by the formula in eq. \ref{U_in_Hge} without the projector $P_{\Hge}$, is a constant depth circuit of local unitaries in $\HgN$.  Its key property is that it has a mixed anomaly with the electric $\Z_N$ one-form symmetry $Z_e^{(1)}$.  Specifically, $U_{\frac{1}{N}}(M^3)$ is an entangler for a $Z_e^{(1)}$ $1$-form SPT. The easiest way to see this is by acting with $U_{\frac{1}{N}}(M^3)$ on a particular short range entangled state in $\HgN$, namely one where $a-d\phi$ has been `Higgsed' to take integer values, and $N$-fold monopoles have been condensed, effectively making $a-d\phi$ $\Z_N$-valued.  Then the expression in eq. \ref{U_in_Hge} is precisely the one-form SPT entangler for a $\Z_N$ one-form degree of freedom written down in ref. \cite{Tsui}.  Another way of diagnosing the mixed anomaly is by localizing the action of $Z_e^{(1)}$ to the boundary of the SPT created by $U_{\frac{1}{N}}(M^3)$, and seeing that these anomalous $Z_e^{(1)}$ line operators are precisely the Wilson lines of the chiral $\Z_N$ anyon TQFT.  In particular, the mixed anomaly implies that $U_{\frac{1}{N}}(M^3)$ is a $Z_e^{(1)}$-protected quantum cellular automaton (QCA) on $\HgN$ \cite{GNVW,gong2020,zhang2023}.  We note that the one-form SPT produced by $U_{\frac{1}{N}}(M^3)$ acting on a trivial product state in $\Hge$ has a non-chiral boundary theory, and it is only the boundary action of the $\Z_e^{(1)}$ symmetry that is chiral.  On the other hand, when acting in $\Hgm$, $U_{\frac{1}{N}}(M^3)$ produces a Walker-Wang wavefunction, which does admit a chiral commuting projector boundary.  However, in contrast to the intrinsic QCA that create such wavefunctions in a tensor product Hilbert space \cite{YuAnQCA2025, HaahQCA}, $U_{\frac{1}{N}}(M^3)$ is not extendable to the tensor product Hilbert space $\Hg$.

The interpretation of the non-invertible chiral symmetry as a one-form SPT entangler under one-form gauging is consistent with its field theory based construction using one-form gauging \cite{Choi_standard_model, Cordova_Ohmori}.  For example, in eq. 2.18 of ref. \cite{Choi_standard_model}, the authors construct the action of the non-invertible chiral symmetry of QED by a twisted gauging of its magnetic $\Z_N$ one-form symmetry:
\begin{align*}
\mathcal{L}_{\text{QED}} \rightarrow \mathcal{L}_{\text{QED}}
+ i\, b \wedge F 
+ i N\, b \wedge d c 
+ i N k\, b \wedge b
\end{align*}
Although our construction applies to a slightly different field theory, one containing only bosonic charges, the interpretation of this field theory approach is closely related to our lattice-based construction: $b$ is the dynamical $2$-form gauge field gauging the $Z_m^{(1)}$ symmetry, and the $i N k\, b \wedge b$ term is the spacetime action for the one-form SPT of the dual $Z_e^{(1)}$ symmetry created by the SPT entangler (here $c$ is a Lagrange multiplier making $b$ a $\mathbb{Z}_N$ gauge field).

The rest of the paper is structured as follows.  In section \ref{sec:U1A} we construct the action of an axial $U(1)_A$ symmetry in a `matter' Hilbert space consisting of rotors, and show that it has the requisite mixed anomaly with the onsite global $U(1)_V$ symmetry.  We also gauge the $U(1)_V$ symmetry by coupling the rotors to a lattice gauge field in the Villain formulation, and construct a gauged (non-conserved) axial charge density $*j_A$.  In section \ref{sec:non-inv-sym} we write down the expression for our non-invertible chiral $\Z_N$ symmetry in $\Hgm \subset \Hg$, constructed to have the same action on the matter fields as $\exp\left[\frac{2\pi i}{N}\int_{M^3} *j_A\right]$.  We also find a dual expression for it by gauging the $\Z_N$ magnetic one-form symmetry $Z_m^{(1)}$.  We show that the resulting operator, acting in $\Hge \subset \HgN$, is an entangler for a $Z_e^{(1)}$ SPT.  In section \ref{sec:FieldTheory} we connect our results to field theory.  We construct a QFT with $U(1)_V$ and $U(1)_A$ global symmetries whose anomaly structure matches that of our lattice rotor model, and give a heuristic argument for why this field theory can emerge from the lattice rotor model.  Upon gauging the $U(1)_V$ in the QFT, the $\Z_N$ remnant of $U(1)_A$ becomes a non-invertible symmetry, and we argue also that this gauged QFT can emerge from our lattice gauge theory model.  We conclude in section \ref{sec:conclusions} with some open directions.

\section{$U(1)_A$ symmetry on the lattice} \label{sec:U1A}

In this section we construct a bosonic lattice Hilbert space of rotors with two $U(1)$ symmetries.  The first will be $U(1)_V$, which will act onsite by just shifting the rotors.  The second will be $U(1)_A$, which will be non-onsite, but generated by a local charge density $*j^{\text{matter}}_A$.  These will have a mixed `t Hooft anomaly, captured by the fact that acting with $U(1)_A$ will pump out non-zero $U(1)_V$ Hall conductivity.  We will also gauge the $U(1)_V$ at the level of the lattice operators, using the Villain formalism, and write down the corresponding gauged local $U(1)_A$ charge density $*j_A$.  We will see that the total charge $\int_{M^3} *j_A$ (where $M^3$ is the spatial manifold) will not commute with the electric field term for the $U(1)_V$ gauge field, as expected from the ABJ anomaly.  Additionally, we will see that $*j_A$ will generate an $\RR$ symmetry instead of $U(1)$.  

However, when restricted to the subspace $\Hmatter \subset \Hg$, $*j_A$ will agree with $*j_A^{\text{matter}}$, and hence generate a bona-fide $U(1)_A$ symmetry.  Our goal in the subsequent section will be to construct a non-invertible $\Z_N$ symmetry whose action, when restricted to this matter subspace, matches the $\Z_N$ subgroup of this $U(1)_A$.  Roughly speaking, this means that the non-invertible $\Z_N$ symmetry, defined in the $U(1)_V$ gauge theory, will act in the correct chiral way on the `matter' degrees of freedom.

\subsection{Matter Hilbert space}

We start with a Hilbert space describing the `matter' degrees of freedom, without any gauge fields:
\begin{align} \label{Hmatter}
{\cal{H}}_{\text{matter}} = \bigotimes_{v} {\cal{H}}^{\text{rotor}}_v
\end{align}
where ${\cal{H}}^{\text{rotor}}_v = L^2(S^1)$, and the tensor product is over vertices $v$ of a simplicial decomposition of a $3$-manifold $M^3$.  We also assume a branching structure on $M^3$, which will be necessary for defining lattice-level cup products \cite{Ellison2021, ChenTata2023}.  We will take the coordinate $\phi_v$ (the angular coordinate normalized by a factor of $\frac{1}{2\pi}$) to be valued in $\RR$, and view states in ${\cal{H}}_{\text{matter}}$ as wavefunctions $\psi(\{\phi_v\})$ invariant under independent integer-valued shifts of $\phi_v$.  Viewing $\{\phi_v\}$ as a $0$-cochain $\phi \in C^0(M^3, \RR)$, this can be expressed as

\begin{align*}
\psi(\phi+s) = \psi(\phi)
\end{align*}
for all $s \in C^0(M^3,\Z)$.  This Hilbert space is equipped with a natural onsite action of a global $U(1)_V$ symmetry that rotates all the rotors simultaneously.  

We now define a non-onsite action of the $U(1)_A$ symmetry by

\begin{align*}
U^{\text{matter}}_\theta(M^3) = \exp\left[i\theta\int_{M^3} *j^{\text{matter}}_A\right]
\end{align*}
with
\begin{align*}
*j^{\text{matter}}_A = (d\phi-\lr{d\phi})\cup d\lr{d\phi}
\end{align*}
Here $\lr{x}$ denotes the integer closest to $x$, so that this is crucially not a continuous function of the field variables. The interpretation of this expression is as follows. $*j^{\text{matter}}_A$ is an $\RR$-valued simplicial $3$-form, i.e. a collection of $\RR$-valued functions for each $3$-simplex of $M^3$.  The integral over $M^3$ is just the oriented sum over all top-dimensional simplices.  Since

\begin{align*}
(d(\phi+s)-\lr{d(\phi+s)})\cup d\lr{d(\phi+s)} = (d\phi-\lr{d\phi})\cup d\lr{d\phi},
\end{align*}
for any $s \in C^1(M^3, \Z)$, each such $\RR$-valued function can be viewed as a well defined Hermitian operator in ${\cal{H}}_{\text{matter}}$, diagonal in the $\phi$ basis.  This makes $U^{\text{matter}}_\theta(M^3)$ into a unitary operator for all $\theta$.  This definition is motivated by the definition of the $2+1$d bosonic SPT partition function in ref.~\cite{DeMarco_SPT}, but in our setting $*j_A^{\text{matter}}$ is an operator in $3+1$d. 

For a closed manifold $M^3$, 
\begin{align} \label{chargedaction}
    U^{\text{matter}}_\theta(M^3) &= \exp\left[i \theta \int_{M^3} \left(d\phi - \lr{d\phi}\right) \cup d\lr{d\phi} \right] \nonumber \\
    &= \exp\left[-i \theta \int_{M^3} \lr{d\phi} \cup d\lr{d\phi}\right]
\end{align}
since $d\phi \,\cup\, d\lr{d\phi}$ is a total derivative.  In particular, this means that $U^{\text{matter}}_{2\pi}(M^3)=1$, so that $U^{\text{matter}}_\theta(M^3)$ does indeed define the action of a $U(1)$ symmetry (as opposed to an $\RR$-symmetry) on a closed $M^3$.

\begin{figure}
    \centering
\includegraphics[width=0.4\linewidth]{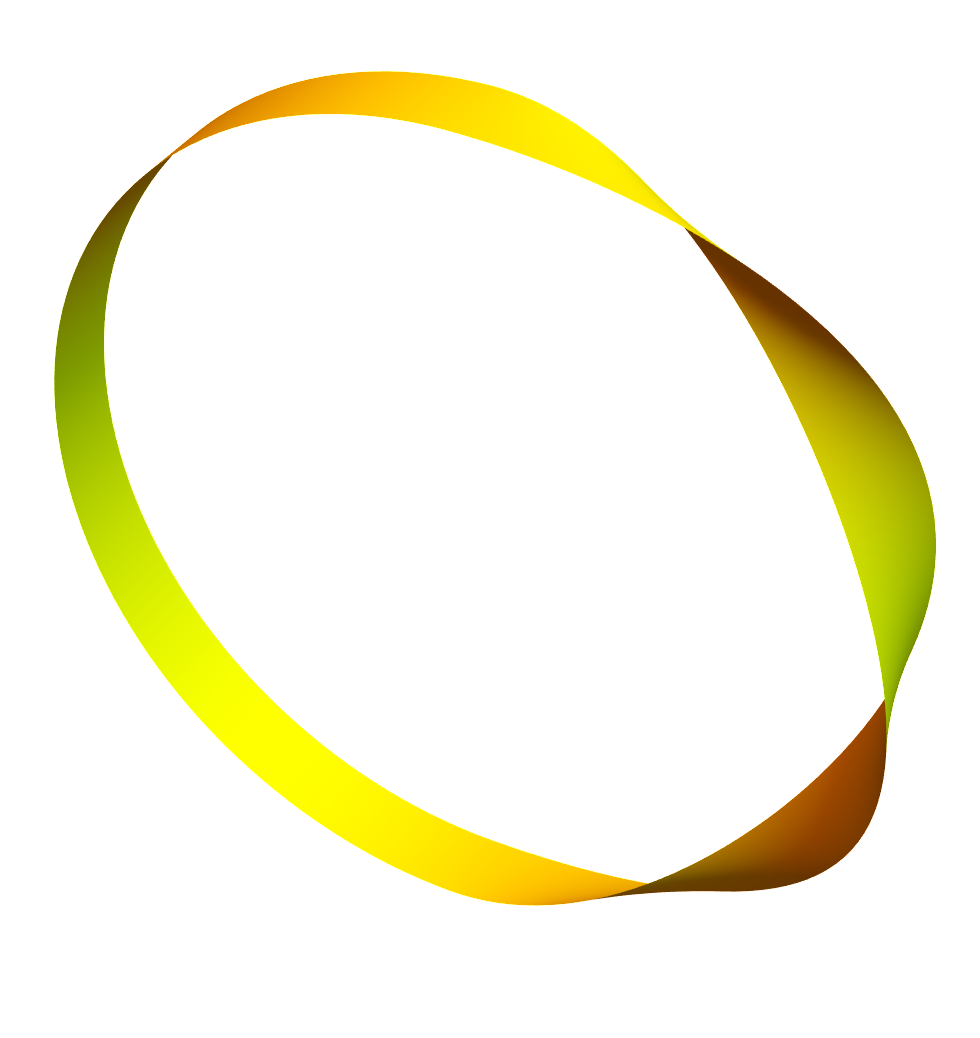}
    \caption{ An example of vortex self-linking. The vortex loop is in fact a ribbon, and Eq.~\ref{chargedaction} gives each linking of the ribbon a phase. }
    \label{vortexlink}
\end{figure}

One interpretation of eq. (\ref{chargedaction}) can be obtained by considering its action on a `superfluid' state, i.e. a single configuration $|\phi\rangle$ consisting of a field $\phi$ which varies slowly everywhere except at the location of branch cuts, where it jumps by $1$.  The $2$ dimensional membrane describing the branch cuts is Poincare dual to $\lr{d\phi}$, and its boundary, Poincare dual to $d\lr{d\phi}$, describes vortex lines in the superfluid.  The exponent in eq. (\ref{chargedaction}) then computes the integer self-linking number of these vortex lines (see Fig.~\ref{vortexlink}).

Now consider a manifold $M^3$ with boundary.  We have
\begin{align*}
U^{\text{matter}}_{2\pi}(M^3) &= \exp\left[2\pi i \int_{M^3} 
 \left(d\phi - \lr{d\phi}\right)\cup d\lr{d\phi}\right] \\ &= \exp\left[-2\pi i \int_{M^3}d(d\phi \cup \lr{d\phi})\right] \\ &= \exp\left[- 2\pi i \int_{\partial M^3}d\phi \cup \lr{d\phi}\right]
\end{align*}
Thus, the action of $U^{\text{matter}}_{2\pi}(M^3)$ on a trivial insulator $|\phi_0\rangle$, namely an equal amplitude superposition of all $\phi$ configurations, is to pump the $2$ dimensional state with wavefunction $\exp\left[- 2\pi i \int_{\partial M^3}d\phi \cup \lr{d\phi}\right]$ out to the boundary of $M^3$.  As shown in \cite{DeMarco_SPT}, this is the root $2$-dimensional SPT state of bosons, with Hall conductivity $2$ in natural units.  This property is a manifestation of the expected mixed anomaly between $U(1)_A$ and $U(1)_V$. Similar construction of 3d bosonic SPT wave functions based on nontrivial vortex loop linking and partition function of 2d SPT states was also discussed in Ref.~\cite{xusenthil1}.

\subsection{Gauging $U(1)_V$}

We now want to examine the fate of $*j^{\text{matter}}_A$ after $U(1)_V$ is gauged.  Of course, the gauged version of $*j^{\text{matter}}_A$, which we denote by $*j_A$, will fail to be conserved, due to the ABJ anomaly.  However, we will want an expression for this non-conserved $*j_A$ as the first step towards constructing our conserved non-invertible symmetry.

In order to perform the gauging at the level of the lattice, we use the Villain formalism.  Thus, in addition to $\phi \in C^0(M^3, \RR)$, we introduce edge degrees of freedom $a \in C^1(M^3, \RR)$ (the $U(1)_V$ vector potential) and Villain variables $n \in C^2(M^3, \Z)$.  Our gauge theory Hilbert space ${\Hg}$ consists of wavefunctions $\Phi(\phi,a,n)$ invariant under the usual gauge invariance
\begin{align} \label{usual_gauge_invariance}
\phi &\rightarrow \phi+\lambda \nonumber \\
a &\rightarrow a + d\lambda
\end{align}
for $\lambda \in C^0(M^3, \RR)$, as well as under the Villain condition
\begin{align} \label{Villain}
a &\rightarrow a + v \\ \nonumber
n &\rightarrow n + dv
\end{align}
for $v \in C^1(M^3, \Z)$.

Before gauging $*j_A$, it will be useful to first define a lattice version of the exponentiated Chern-Simons functional $\exp\left[2 \pi i \int_{M^3} a \, da\right]$.  Following \cite{Jacobson} we do this by replacing
\begin{align} \label{lattice_CS}
a\, da \rightarrow da \cup (a-d\phi) - (a-d\phi)\cup n - n\cup (a-d\phi) - (a-d\phi) \cup_1 dn
\end{align}
where the higher cup product $\cup_1$ is defined in \cite{Jacobson}.  As shown in \cite{Jacobson}, the expression on the right hand side of eq. \ref{lattice_CS} changes by an integer under the Villain transformation (eq. \ref{Villain}).  It is also manifestly invariant under the usual gauge invariance (eq. \ref{usual_gauge_invariance}), so that multiplying by $2\pi i$ and taking the exponential yields a well-defined local operator.  Furthermore, as shown in \cite{Jacobson}, for a closed $M^3$ we have
\begin{align*}
\exp & \left[2\pi i\int_{M^3}\left(da \cup (a-d\phi) - (a-d\phi)\cup n - n\cup (a-d\phi) - (a-d\phi) \cup_1 dn\right)\right] \\ &= \exp \left[2\pi i\int_{M^3}\left(da \cup a - a\cup n - n\cup a - a \cup_1 dn - 2\phi\cup dn\right)\right],
\end{align*}
so that in particular this expression is independent of the matter field $\phi$ on configurations with no monopoles ($dn=0$).

We will now write down a local Hermitian operator $*j_A$ in ${\Hg}$ which will perform multiplication by a function of $\phi, a, n$; this function will also be denoted $*j_A$.  In order to define a valid operator on $\Hg$, $*j_A$ will have to be invariant under eq. (\ref{usual_gauge_invariance}) and (\ref{Villain}).  $*j_A$ will also have the following two properties. First, it will agree with $*j_A^{\text{matter}}$ on the configurations $\{\phi, a,n\}=\{\phi, 0,0\}$, so that it may be thought of as a gauging of $*j_A^{\text{matter}}$.  Second, it will satisfy, for a closed $M^3$:

\begin{align} \label{exp_condition}
\exp & \left[2\pi i \int_{M^3} *j_A\right] = \nonumber \\ &\exp \left[2\pi i\int_{M^3}\left(da \cup a - a\cup n - n\cup a - a \cup_1 dn - 2\phi\cup dn\right)\right].
\end{align}
This may be thought of as the lattice level analogue of the continuum equation $\exp\left[2\pi i\int_{M^3}(*j_A-a\,da)\right] = 1$, which follows in the continuum from $d*j_A = da \wedge da$.  Note that in particular this implies that $*j_A$ generates only an $\RR$ symmetry, since the integral of $a\,da$, or its lattice version in eq. \ref{lattice_CS}, can be any real number.

A function $*j_A$ that satisfies all these properties is:
\begin{align} \label{jAtotal}
*j_A = *j_A^{(I)} + *j_A^{(II)}
\end{align}
where 
\begin{align} \label{jAI}
*j_A^{(I)}&=da \cup (a-d\phi) - (a-d\phi)\cup n - n\cup (a- d\phi) - (a-d\phi) \cup_1 dn
\end{align}
and
\begin{align} \label{jAII}
*j_A^{(II)} = &- \lfloor a-d\phi \rceil\cup d \lfloor a-d\phi \rceil + \lfloor a-d\phi \rceil \cup n + n \cup \lfloor a-d\phi \rceil \nonumber \\ &+ \lfloor a-d\phi \rceil \cup_1 dn - d((a-d\phi) \cup \lfloor a-d\phi \rceil)
\end{align}
Note that we included a total derivative in the definition of $*j_A^{(II)}$ in order to make $*j_A$, and not just its integral, invariant under the Villain condition (eq. \ref{Villain}). However, $*j_A^{(I)}$ and $*j_A^{(II)}$ are not independently well defined operators, because they are not individually invariant under eq. \ref{Villain}.  Also, because $*j_A^{(II)}$ is an integer, the condition in eq. \ref{exp_condition} is satisfied.  Furthermore, $*j_A^{(I)}(\phi, 0, 0) = 0$ and $*j_A^{(II)}(\phi,0,0) = d\phi \cup d\lr{d\phi}-\lr{d\phi} \cup d\lr{d\phi}$, so that $*j_A(\phi,0,0) = *j_A^{\text{matter}}(\phi)$.  Because of Villain and gauge invariance, we have more generally that $*j_A(\phi+\lambda,d\lambda + v, dv) = *j_A^{\text{matter}}(\phi)$, for $\lambda \in C^0(M^3,\RR)$ and $v \in C^1(M^3,\Z)$.  Wave functions supported on these configurations define the subspace $\Hmatter \subset \Hg$, so we have that on $\Hmatter$, and closed $M^3$:
\begin{align}
\begin{split}\label{flat}
&P_{\mathcal{H}_{\mathrm{matter}}}\left(\exp\left[\frac{2\pi i}{N}\int_{M^3} *j_A\right]\right)P_{\mathcal{H}_{\mathrm{matter}}}\\ & = P_{\mathcal{H}_{\mathrm{matter}}}\left(\exp \left[-\frac{2\pi i}{N} \int_{M^3} \lr{a-d\phi}\cup d\lr{a-d\phi}\right]\right)P_{\mathcal{H}_{\mathrm{matter}}}.
\end{split}\end{align}

\section{Non-invertible chiral $\Z_N$ symmetry} \label{sec:non-inv-sym}

Let us now write down a non-invertible chiral $\Z_N$ symmetry operator $U_{\frac{1}{N}}(M^3)$ that reduces to eq. \ref{flat} on flat gauge field configurations.  The non-invertible nature of $U_{\frac{1}{N}}(M^3)$ will come from a projection $P_{\Hgm}$ onto the subspace $\Hgm$ of states invariant under the $\Z_N$ magnetic one-form symmetry $Z_m^{(1)}$.  Specifically, for any homology $2$-surface $\Sigma \in M^3$, we define
\begin{align*}
Z_m^{(1)}(\Sigma)\cdot \Psi(\phi,a,n) = \exp\left[\frac{2 \pi i}{N}\int_\Sigma n\right] \Psi(\phi,a,n),
\end{align*}
and define
\begin{align*}
\Hgm = \{|\psi\rangle \in \Hg \,| \,Z_m^{(1)}(\Sigma)\cdot|\psi\rangle = |\psi\rangle \text{ for all } \Sigma\}
\end{align*}
We note that while $\Hgm$ does not contain configurations with single magnetic monopoles, it does allow $N$-fold monopoles.  The constraint that $\exp\left[\frac{2 \pi i}{N}\int_\Sigma n\right] = 1$ for all $\Sigma$ means that $n \text{ mod } N$ is an exact co-cycle in $C^2(M^3, \Z_N)$, which in turn implies that $n = dv + Nn'$ where $v \in C^1(M^3,\Z)$ and $n' \in C^2(M^3, \Z)$.  Let us define $d^{-1}n \equiv v$.  The ambiguity in this definition is $d^{-1}n \rightarrow d^{-1}n + \alpha$, where $\alpha$ is any co-chain in  $C^1(M^3, \Z)$ that satisfies $d\alpha \in C^2(M^3,N\Z)$. 
 We then define, in $\Hgm \subset \Hg$,
\begin{align} \label{noninv_sym_expression}
U_{\frac{1}{N}}&(M^3)\cdot \Psi(\phi,a,n) \nonumber \\ &= \exp\left[-\frac{2\pi i}{N} \int_{M^3} \left(\lr{a-d\phi} - d^{-1}n\right)\cup\left(d \lr{a-d\phi} -n \right)\right]\cdot\Psi(\phi,a,n)
\end{align}
To see that this operator $U_{\frac{1}{N}}(M^3)$ is well defined, we first note that it is invariant under $d^{-1}n \rightarrow d^{-1}n + \alpha$, with $\alpha$ as above.  Also, it is manifestly gauge invariant (eq. \ref{usual_gauge_invariance}), and invariant under the Villain transformation (eq. \ref{Villain}).  Additionally, it reduces to $\exp\left[\frac{2\pi i}{N}\int_{M^3} *j_A\right]$, as seen from eq. \ref{flat}, on $\Hmatter$.  However, from this perspective the locality properties of $U_{\frac{1}{N}}(M^3)$ are not entirely manifest, due to the presence of the $d^{-1}n$ term.  In the following we will give a slightly different derivation of eq. \ref{noninv_sym_expression} which will make its locality properties more manifest, and make the connection to field theory more clear.

\subsection{Gauging magnetic $\Z_N$ one-form symmetry}

We gauge the $Z_m^{(1)}$ one-form symmetry by coupling to a pure-gauge $2$-form dynamical $\Z_N$ gauge field.  Just as the Kramers-Wannier duality in $1+1$d can be thought of as resulting from gauging the $\Z_2$ $0$-form symmetry, gauging $Z^{(1)}_m$ in our $3+1$d Hilbert space $\Hg$ will result in a dual electric 1-form symmetry $Z_e^{(1)}$ acting in $\HgN$, the Villain lattice gauge theory of a charge $\frac{1}{N}$ scalar.  Defining $\Hge$ as the subspace of $\HgN$ invariant under $Z_e^{(1)}$, we then have that the symmetric subspaces $\Hgm$ and $\Hge$ are isomorphic.
While the $U_{\frac{1}{N}}(M^3)$ constructed in eq. \ref{noninv_sym_expression} does not extend as a locality-preserving unitary operator to all of $\Hg$ - i.e. it is non-invertible on $\Hg$ - we will see that, when viewed as an operator on $\Hge$, it does extend as a local quantum circuit to all of $\HgN$, albeit one with a mixed anomaly with $Z_e^{(1)}$.

We define $\HgN$ to consist of wavefunctions $\Psi(\phi,a,n)$, with $\phi \in C^0(M^3, \RR)$, $a \in C^1(M^3,\RR)$, and $n\in C^2(M^3, N\Z)$, invariant under the usual gauge invariance $\phi \rightarrow \phi + \lambda, a\rightarrow a+ d\lambda$ for $\lambda \in C^0(M^3, \RR)$, as well as the Villain condition rescaled by a factor of $N$
\begin{align} \label{mod_Villain}
a &\rightarrow a + Nv \\ \nonumber
n &\rightarrow n + Ndv
\end{align}
for $v \in C^1(M^3, \Z)$.  $\HgN$ is simply the Villain lattice gauge theory Hilbert space of a charge $\frac{1}{N}$ scalar.

We then define the action of an `electric' $\Z_N$ one-form symmetry $Z^{(1)}_e$ in $\HgN$ as follows.  Let $v \in C^1(M^3, \Z)$ be a cochain that satisfies $dv \in C^2(M^3, N\Z)$.  Such a $v$ defines a cocycle in $C^1(M^3,\Z_N)$ under a reduction of coefficients mod $N$, which by abuse of notation we will also denote $v$.  We then define $Z^{(1)}_e(v)$ in terms of its action on wavefunctions as follows:
\begin{align} \label{ZNe}
Z^{(1)}_e(v): \Psi(\phi,a,n) \rightarrow \Psi(\phi, a-v, n-dv)
\end{align}
Let $\Hge$ be the $Z^{(1)}_e$-invariant subspace of $\HgN$:
\begin{align*}
\Hge = \{|\psi\rangle \in \HgN \,| \,Z_e^{(1)}(v)\cdot|\psi\rangle = |\psi\rangle \text{ for all } v\in C^1(M^3,\Z) \text{ s.t. } dv\in N\Z\}
\end{align*}
Physically, $\Hge$ is a Hilbert space with integer electric charge Wilson lines and $N$ times integer magnetic charge `t Hooft lines.  In particular, combining the usual gauge invariance (eq. \ref{usual_gauge_invariance}) with integer valued $\lambda$ together with $Z^{(1)}_e(d\lambda)$ shows that all states in $\Hge$ are invariant under $\phi_v \rightarrow \phi_v + 1$.

To see how $\Hge \subset \HgN$ is the result of gauging the $Z_m^{(1)}$ one-form symmetry of $\Hgm$ by minimally coupling to a $2$-form dynamical $\Z_N$ gauge field at the level of the Hamiltonian lattice, see the derivation in appendix \ref{app:gauging_appendix}.  Intuitively, we expect that under such $\Z_N$ gauging, the $\Z_N$ degrees of freedom should shift from living on the faces to living on the edges (faces of the dual lattice), in the same way that in 1+1d Kramers-Wannier duality the $\Z_2$ degrees of freedom shift from the original to the dual lattice.  This is indeed what happens when we go from $\Hgm$ to $\Hge$: the Villain variable $n$, defined on faces, goes from being integer valued to being $N$ times integer valued, while the spacing between the allowed quantized electric field values on edges goes from $1$ in $\Hgm$ to $\frac{1}{N}$ in $\Hge$.  This $N$-fold increase in the allowed values for the electric field can be viewed as the introduction of a $\Z_N$ dual $2$-form gauge field.  For more details see appendix \ref{app:gauging_appendix}.  

It is worth emphasizing that the unitary duality isomorphism 
\begin{align} \label{def_rho}
\rho:\Hgm \xrightarrow{\cong} \Hge
\end{align}
resulting from this one-form gauging is a map between two different Hilbert spaces.  The analogous map in the $1+1$d $\Z_2$ setting would be
\begin{align*}
X_i &\rightarrow X_{i-1/2} X_{i+1/2} \\
Z_i Z_{i+1} &\rightarrow Z_{i+1/2}
\end{align*}
This is the map that results from coupling a system of $\Z_2$ `matter' spins (living on integer sites) to a $\Z_2$ gauge field (living on integer plus one-half sites), imposing the Gauss law, and reading off the wavefunction in a gauge where the matter spins all point up.  Its domain is the Hilbert space of matter spin wavefunctions invariant under the spin flip $\prod_i X_i$ symmetry, and its range is the space of gauge field wavefunctions invariant under $\prod_i Z_{i+1/2}$.  The latter is just the holonomy of the $\Z_2$ gauge field, which is conserved because the $\Z_2$ gauge field has no dynamics.  To make contact with the usual Kramers-Wannier transformation one composes with a Hadamard transformation ($X \leftrightarrow Z$) and a half-integer spatial shift, which maps $\prod_i Z_{i+1/2}$ back to $\prod_i X_i$ and hence results in a map from the original Hilbert space to itself.  In our $3+1$d situation, we will not be performing this second step; in fact, it will be important for us that our duality isomorphism $\rho$ is a map between two different presentations of the symmetric Hilbert space.

At the level of wavefunctions, our unitary duality isomorphism $\rho$ can be described by
\begin{align*}
\rho(\Psi)(\phi,a,n) = c \, \Psi(\phi,a,n)
\end{align*}
where
\begin{align*}
c = \frac{N^{(|\text{faces}|-|\text{edges}|)/2}}{\sqrt{|{\{\beta|\partial\beta=0\}|}}}
\end{align*}
and $\partial = \dd$ is the Hodge dual of the coboundary operator.  This is derived in appendix \ref{subsec:HgmHge} (see eq. \ref{derivedrho}).  In other words, given a wavefunction $\Psi$ in $\Hgm$, up to a normalization constant the corresponding wavefunction $\rho(\Psi) \in \Hge$ is simply given by restricting $\Psi$ to configurations where $n\in C^2(M^3, N\Z)$. Note that due to invariance under the Villain condition, this restriction never annihilates the state, ensuring that the isomorphism is indeed unitary. To describe $\rho^{-1}$, suppose $\Psi' \in \Hge$, and suppose that $n = dv + Nn'$, for $v\in C^1(M^3,\Z)$, $n' \in C^2(M^3,\Z)$.  Then
\begin{align} \label{rho_inverse}
\rho^{-1}(\Psi')(\phi,a,n) = c^{-1}\, \Psi'(\phi,a-v,Nn')
\end{align}
The fact that the above expression is independent of $v$ follows from the $\Z_e^{(1)}$-invariance of $\Psi'$.

The isomorphism $\rho$ between $\Hgm$ and $\Hge$ can also be analyzed at the level of its action on local operators.  To find the local operator algebras of these two Hilbert spaces, first let us consider the local operator algebra ${\cal{A}}(\Hg)$.  This is generated by the electric field operator $e_l = \frac{1}{2\pi i} \cdot \frac{d}{da_l}$ for every edge $l$, the magnetic field operator $b_f = (da-n)_f$ for every face $f$, the open Wilson line operator $W_l = \exp(2\pi i (a-d\phi)_l)$ for every edge $l$, and the open `t Hooft line operator $T_{m}: \Psi(\phi,a,n) \rightarrow \Psi(\phi,a,n+m)$ for any $n'\in C^2(M^3,\Z)$:
\begin{align*}
{\cal{A}}(\Hg) = \langle \{e_l\}_l, \{b_f\}_f,\{W_l\}_l, \{T_{m}\}_{m \in C^2(M^3,\Z)} \rangle.
\end{align*}
Now, wavefunctions in $\Hgm$ are supported on configurations $n = dv +Nn'$, $v\in C^1(M^3,\Z)$, $n'\in C^2(M^3, \Z)$, so the operator algebra is generated by the same operators, except the `t Hooft lines can only shift $n$ in multiples of $N$:
\begin{align*}
{\cal{A}}(\Hgm) = \langle \{e_l\}_l, \{b_f\}_f,\{W_l\}_l, \{T_{m}\}_{m \in C^2(M^3,N\Z)} \rangle.
\end{align*}
By the same argument, $\HgN$ has Wilson lines that transport fractional $\frac{1}{N}$ charge, but $\Hge\subset \HgN$ has only integer Wilson line operators:
\begin{align*}
{\cal{A}}(\Hge) = \langle \{e_l\}_l, \{b_f\}_f,\{W_l\}_l, \{T_{m}\}_{m \in C^2(M^3,N\Z)} \rangle.
\end{align*}
Now, the isomorphism from ${\cal{A}}(\Hgm)$ to ${\cal{A}}(\Hge)$ induced by $\rho$ is simply the one that looks like the identity in the above presentation.  Indeed, this is clear from the wavefunction formula for $\rho$ (eq. \ref{def_rho}).

\subsection{Mixed anomaly between $U_{\frac{1}{N}}(M^3)$ and $Z^{(1)}_e$}

The non-invertible chiral $\Z_N$ symmetry $U_{\frac{1}{N}}(M^3)$ is defined in eq. \ref{noninv_sym_expression}.  It is a unitary operator within $\Hgm\subset\Hg$.  Conjugating by the isomorphism $\rho$ in eq. \ref{def_rho}, one obtains the following operator, which by abuse of notation we also refer to as $U_{\frac{1}{N}}(M^3)$:
\begin{align} \label{tentative_def}
U_{\frac{1}{N}}(M^3) = \exp\left[-\frac{2\pi i}{N}\int_{M^3} \lr{a-d\phi}\cup d\lr{a-d\phi}\right]
\end{align}
Note that $U_{\frac{1}{N}}(M^3)$ is manifestly a constant depth quantum circuit of local unitaries in $\HgN$, commutes as an operator with $Z^{(1)}_e$, and hence defines a unitary operator in $\Hge$.

The non-trivial property of $U_{\frac{1}{N}}(M^3)$ is that it has a mixed anomaly with $Z_e^{(1)}$.  One way to diagnose this mixed anomaly is as follows.  First, define $|\psi_{\text{Higgs}}\rangle \in \HgN$ to be an equal-amplitude superposition of all configurations $\{\phi,a,n\}$ for which $r \equiv a-d\phi \in C^1(M^3, \Z)$.  Since this constraint pins the covariant derivative of $\phi$, it a Higgsed state.  Then $U_{\frac{1}{N}}(M^3)|\psi_{\text{Higgs}}\rangle$ is a state whose amplitude depends only on $r = a-d\phi$ (in particular, not on $n$), and is $\exp\left[-\frac{2\pi i}{N} \int_{M^3} r \cup dr\right]$.  This is the wavefunction of a non-trivial one-form SPT of $Z^{(1)}_e$ \cite{Tsui}.  This means that although $U_{\frac{1}{N}}(M^3)$ is a constant depth circuit, its gates cannot all be invariant under $Z^{(1)}_e$, so that $U_{\frac{1}{N}}(M^3)$ is a symmetry-protected QCA like all (symmetric) SPT entanglers.  This is equivalent to the statement that there is a mixed anomaly between $U_{\frac{1}{N}}(M^3)$ and $Z^{(1)}_e$.  Another way to see this anomaly is to follow the idea of Else and Nayak \cite{Else_Nayak} to construct, the $2+1$d `boundary' action of $Z^{(1)}_e$ at the boundary of the SPT state created by $U_{\frac{1}{N}}(M^3)$, and read off the anomaly from the non-trivial commutation (braiding) relations of the resulting $Z^{(1)}_e$ string operators.  We give more details on this approach in appendix \ref{app:mixed_anomaly_appendix}.

\section{Field Theories with the non-invertible chiral symmetry} \label{sec:FieldTheory}

The main goal of this work was to construct a non-invertible $\Z_N$ chiral symmetry as an operator acting in a many-body lattice gauge theory Hilbert space (eq. \ref{noninv_sym_expression}).  One can then consider lattice Hamiltonians that commute with this non-invertible symmetry.  It is desirable to construct at least one such Hamiltonian that flows, at low energies, to some familiar field theory with a non-invertible chiral $\Z_N$ symmetry.  In this section we first construct such a `target' field theory and discuss some of its key properties, and then argue that it can emerge from our microscopic setup.

\subsection{Target field theory} \label{subsec:target}

The simplest field theory discussed in the context of the non-invertible chiral symmetry is $3+1$ d QED, with $N_f=1$ flavor of Dirac fermion:
\begin{align*}
{\cal L}_{\text{QED}}(\Psi, \bar{\Psi}, A) = \frac{1}{4e^2} F_{\mu\nu} F^{\mu\nu} + i\bar{\Psi}(\partial_\mu-iA_\mu)\gamma^\mu \Psi
\end{align*}
However, one impediment to emerging QED out of a microscopic lattice Hilbert space is that the axial current
\begin{align*}
{j_{\text{QED}}}_A^\mu = \bar{\Psi}\gamma_5\gamma^\mu\Psi
\end{align*}
has, in addition to the standard $U(1)_V$ piece, a gravitational contribution to its divergence ${\cal{D}}_\mu {j_{\text{QED}}}^\mu_A \propto \text{Tr } R\wedge R$ \cite{AlvarezGaume}, where $R$ is the curvature of the Levi-Civita connection.  This turns out to be an obstacle for simple lattice constructions \cite{FidkowskiXu}, so it will be useful to cancel out this gravitational contribution.  For QED, the simplest way to do this is to introduce an additional Dirac fermion $\Psi'$, neutral under the $U(1)_V$, resulting in the action
\begin{align} \label{fermL}
{\cal L}_{\text{QED'}}(\Psi, \bar{\Psi}, \Psi', \bar{\Psi}', A) = \frac{1}{4e^2} F_{\mu\nu} F^{\mu\nu} + i\bar{\Psi}(\partial_\mu-iA_\mu)\gamma^\mu \Psi + i\bar{\Psi}' \partial_\mu \gamma^\mu \Psi'
\end{align}
Because $\Psi'$ still couples to the metric, the gravitational contribution to the anomalous divergence of 
\begin{align*}
{j_{\text{QED'}}}_A^\mu = \bar{\Psi}\gamma_5\gamma_\mu\Psi - \bar{\Psi}'\gamma_5\gamma_\mu\Psi'
\end{align*}
vanishes.  
Then the gravitational contribution to the divergence of ${j_{\text{QED'}}}_A^\mu$ is $0$, whereas the $U(1)_V$ contribution remains non-zero.

One important feature of ${j_{\text{QED'}}}_A^\mu$ is the following.  Let us work in the `un-gauged' theory, where $U(1)_V$ is a global symmetry.  Then we just have a $U(1)_V$-charged Dirac fermion $\Psi$ together with a neutral one $\Psi'$.  Let $B \subset \RR^3$ be a $3$-dimension spatial ball, and define $Q_A^{(B)} = \int_{B} {j_{\text{QED'}}}_A^0$ to be the operator that measures the axial charge in $B$.  Also, let $|\Psi_0\rangle$ be the short-range entangled state obtained by gapping out both $\Psi$ and $\Psi'$.  Then $\exp (2\pi i Q_A^{(B)})|\Psi_0\rangle$ looks like $|\Psi_0\rangle$ away from the boundary of $B$\footnote{This follows from the fact that $\exp (2\pi i Q_A^{(B)})|\Psi_0\rangle=\exp (2\pi i Q_A^{(\bar{B})})|\Psi_0\rangle$ where $Q_A^{(\bar{B})}$ is the axial charge in the complement of $B$.}, whereas near the boundary it has an IQH state for the $\Psi$ fermions (electric Hall conductance $\frac{e^2}{h}$ and chiral central charge $1$) stacked with an IQH state for the $\Psi'$ fermions (electric Hall conductance $0$ and chiral central charge $1$).  This is precisely the root $2+1$d fermionic symmetry protected topological (SPT) state for $U(1)$ symmetry, which carries a Hall conductivity of $\frac{e^2}{h}$.  Thus the current ${j_{\text{QED'}}}_A^\mu$ generates a unitary that, at rotation angle $2\pi$, `pumps out' the root $U(1)$ SPT state of fermions.  This is the key property that encodes the mixed `t Hooft anomaly between $U(1)_V$ and $U(1)_A$.

In this work we will target a field theory that has a bosonic analogue of this property: namely, the axial current generates a unitary that `pumps out' the root $2+1$d $U(1)$ SPT state of bosons, with Hall conductivity $\frac{2e^2}{h}$ \cite{LevinSenthil}.  To find the target field theory with this property, we must simply find a bosonic version of (\ref{fermL}), i.e. a theory that can emerge out of an underlying bosonic Hilbert space, as opposed to a $\Z_2$-graded one.

Since we want a theory that `pumps out' the root $U(1)$ SPT state of bosons, let us first discuss a parton construction for such $2+1$d bosonic SPT. Let us consider fermionic partons $\psi_1$, $\psi_2$ and $f_1$, $f_2$.
We assign $\psi_\alpha$ physical charge 1, and $f_\alpha$ physical charge 0. $\psi_\alpha$ and $f_\alpha$ are also both SU(2) fundamental doublets of a SU(2) gauge group. 
All physical objects should be gauge invariant bosons with integer charges. One gauge invariant boson is \begin{align*}
b \sim \psi_1 f_1^\dagger + \psi_2 f_2^\dagger.
\end{align*} 
As a ``mean field" state, we put $\psi_\alpha$ both in a $C = 1$ Chern band, and $f_\alpha$ both in a $C = -1$ Chern band. We claim that this parton state, combined with the SU(2) gauge field, gives us the desired boson SPT state, for the following reasons:

(1) At the mean field level the 1+1d boundary of the 2+1d state constructed above will be a nonchiral CFT $\U(2)_{1}\times \overline{\U(2)}_{1}$, with nonchiral central charge $c = 2$. The boundary CFT decomposes as 
\begin{align*}
\U(2)_1 = \left( \SU(2)_1 \times \U(1)_2 \right) / \mathbb{Z}_2,
\end{align*}
with the same decomposition for $\overline{\U(2)}_{1}$. The SU(2) gauge field will erase the $\SU(2)_1 \times \overline{\SU(2)}_1$ part of the boundary CFT, leaving only a nonchiral $c = 1$ CFT $\U(1)_{2} \times \overline{\U(1)}_{2}$. 

(2) Since $\psi_\alpha $ both carry charge 1, the total Hall conductivity is $\sigma_{xy} = 2$, consistent with our expectation for a bosonic $U(1)$ SPT state.

(3) There is no net Chern-Simons term in the 2+1d for the SU(2) gauge field, as the Chern-Simons terms generated from $\psi_\alpha$ and $f_\alpha$ cancel out. Therefore the SU(2) gauge field can safely confine. 

Now we are ready to construct our target $(3+1)$-dimensional theory.  Let us consider QCD with an $\SU(2)$ gauge field, and $N_f = 2$ flavors of Dirac fermions, one flavor being the $SU(2)$ fundamental $\psi_\alpha$, and having charge $1$ under a global $U(1)_V$ symmetry, and the other being another $SU(2)$ fundamental $f_\alpha$, having charge $0$ under the global $U(1)_V$ symmetry (note that $\alpha$ is the SU(2) color index).  The Lagrangian is
\begin{align} \label{SU2Lagrangian_first}
{\cal{L}}= -\frac{1}{4} \text{Tr}\, G_{\mu\nu} G^{\mu\nu} + \bar{\psi} \,i \gamma^\mu D_\mu \psi + \bar{f}\, i \gamma^\mu D_\mu f,
\end{align}
where the $G_{\mu\nu}$ is the $SU(2)$ field strength and $D_\mu$ is the $SU(2)$ covariant derivative.  We now define the axial rotation to be the usual axial symmetry for $\psi$ combined with the opposite one for $f$.  That is, the Weyl fermions $\psi_L$ and $f_R$ carry axial charge $+1$, and $\psi_R$ and $f_L$ carry axial charge $-1$.  We claim that the $2\pi$ axial rotation will then generate the desired $2d$ SPT state at the boundary, for the following reasons:

(1) If we couple the theory in eq. \ref{SU2Lagrangian_first} to a dynamical $U(1)_V$ gauge field $A_\mu$, the $2\pi$ axial rotation will generate a $\Theta = 4\pi$ term for $A_\mu$ in the $(3+1)d$ bulk, which is consistent with the response of $\sigma_{xy} = \frac{2q^2}{h}$ for the root bosonic $U(1)$ SPT at the boundary. 

(2) The $2\pi$ axial rotation does not generate a net $\Theta$-term for the $SU(2)$ gauge field.

(3) The $2\pi$ axial rotation does not generate a net gravitational $\Theta$-term.

We will now give a heuristic argument that the field theory in eq. \ref{SU2Lagrangian_first} can emerge from our microscopic rotor model while preserving $U(1)_A$ and $U(1)_V$, and that its $U(1)_V$-gauged version can emerge from the corresponding lattice field theory while preserving the non-invertible $\Z_N$ symmetry.

\subsection{Global $U(1)_V$ symmetry}

Let us now give a heuristic argument that the field theory in eq. \ref{SU2Lagrangian_first} can emerge from our microscopic rotor model with global $U(1)_V$ symmetry.  This argument will rest on one crucial assumption: that we can realize two conjugate copies of ${\cal{L}}$ with a microscopic Hamiltonian invariant under onsite actions of both the diagonal $U(1)_A$ and $U(1)_V$.

The Hilbert space ${\cal{H}}_{\text{matter}}$ consists of rotor degrees of freedom on vertices, as defined in eq. \ref{Hmatter}.  The $U(1)_V$ is just the onsite rotation of the rotors, whereas the $U(1)_A$ is generated by
\begin{align*}
*j_A^{\text{matter}} = (d\phi - \lr{d\phi})\cup d\lr{d\phi}
\end{align*}
Our goal is to find a microscopic Hamiltonian on this Hilbert space, together with ancillas described below, which commutes with $U(1)_V$ and $U(1)_A$, and whose low energy dynamics is described by the Lagrangian
\begin{align} \label{SU2Lagrangian}
{\cal{L}}= -\frac{1}{4} \text{Tr}\, G_{\mu\nu} G^{\mu\nu} + \bar{\psi}\, i \gamma^\mu D_\mu \psi + \bar{f}\, i \gamma^\mu D_\mu f,
\end{align}
such that $\psi$ carries charge $1$ under the $U(1)_V$, $f$ is neutral, and the $U(1)_A$ acts according to
\begin{align*}
j^\mu_A = \bar{\psi}\, \gamma_5 \gamma^\mu \psi - \bar{f} \gamma_5 \gamma^\mu f.
\end{align*}
Here $G_{\mu\nu}$ is the $SU(2)$ field strength, and $\psi$ and $f$ are both fundamentals under the $SU(2)$.  We will show that we can do this if we are allowed to introduce ancilla degrees of freedom into the Hilbert space, which transform in an onsite way under both $U(1)_V$ and $U(1)_A$.

To proceed, we first argue that one can realize two conjugate copies, ${\cal{L}}$ and $\cal{L}'$ in an ancilla microscopic Hilbert space ${\cal{H}}_{\cal{L} + \cal{L}'}$ where the diagonal $U(1)_V$ and $U(1)_A$ both act onsite.  By conjugate copy $\cal{L}'$, we mean the same Lagrangian as ${\cal{L}}$:
\begin{align} \label{SU2Lagrangian_prime}
{\cal{L}}'= -\frac{1}{4} \text{Tr}\, G'_{\mu\nu} G'^{\mu\nu} + \bar{\psi'}\, i \gamma^\mu D_\mu \psi' + \bar{f'}\, i \gamma^\mu D_\mu f',
\end{align}
but with $\psi'$ having charge $-1$ under the $U(1)_V$, and $f'$ being neutral, and the $U(1)_A$ acting according to
\begin{align*}
j'^\mu_A = -\bar{\psi'}\, \gamma_5 \gamma^\mu \psi + \bar{f'} \gamma_5 \gamma^\mu f'.
\end{align*}
This is plausible, because with two conjugate copies, the diagonal $U(1)_V$ and $U(1)_A$ symmetries have no net 't Hooft anomalies, so there is no anomaly obstruction to this happening.\footnote{Indeed, if the theory we were dealing with was instead a single Dirac fermion, then it is clear that two conjugate copies of it could be realized in a microscopic Hilbert space with the diagonal $U(1)_V$ and $U(1)_A$ both acting onsite.  This is because the field content of such a doubled theory would have $4$ left handed Weyl fermions, with $U(1)_V$ charges $+1,-1,-1,+1$ and $U(1)_A$ charges $+1,+1,-1,-1$.  These can be paired up into Dirac fermions ($1$ with $3$ and $2$ with $4$) such that both symmetries act in a vector-like way.}  In our theory of interest there are also $SU(2)$ gauge fields, and this kind of re-arrangement would result in chiral, albeit non-anomalous, $SU(2)$ gauge theories.  Although we cannot prove that this is possible, we will still assume that a microscopic Hamiltonian $H_{{\cal{L}}+\cal{L}'}$ acting in ${\cal{H}}_{\cal{L} + \cal{L}'}$ and commuting with an onsite action of $U(1)_V$ and $U(1)_A$ can realize the low energy field theory ${\cal{L}} + {\cal{L}}'$.

The next step is to work in the Hilbert space ${\cal{H}}_{\cal{L} + \cal{L}'} \otimes {\cal{H}}_{\text{matter}}$, starting with the Hamiltonian $H_{{\cal{L}}+\cal{L}'}$ (extending it to act trivially in ${\cal{H}}_{\text{matter}}$).  For specificity, we will assume that, possibly after some coarse-graining, the microscopic spin degrees of freedom of ${\cal{H}}_{\cal{L} + \cal{L}'}$ live on the same vertices $v$ as the matter degrees of freedom.  Our goal is then to deform this Hamiltonian by coupling the matter and ancilla degrees of freedom, thereby gapping out all of the matter and $\cal{L}'$, while preserving the combined $U(1)_V$ and $U(1)_A$.  The only low energy degrees of freedom then left will the be those of ${\cal{L}}$, giving us our desired realization of ${\cal{L}}$. The procedure is sketched below: 

\begin{equation*}
\cH_{\cL + \cL'} \times \cH_{{\rm matter}} \ \ \Rightarrow \ \ \cH_{\cL} \times \left( \cH_{\cL'} \times \cH_{\rm matter} \right)_{\rm Gap} \end{equation*} 

The simplest gapped state is at expense of breaking the $U(1)_A$, by introducing Dirac mass terms for $\cal{L}'$, and the trivial rotor Hamiltonian that puts all the rotors in their $0$ angular momentum state; let us call such a $U(1)_V$-conserving but $U(1)_A$-breaking perturbation $V$:
\begin{align*}
V = \int_{\RR^3}m_0 \left(\psi'^\dagger\gamma^0\psi'+f'^\dagger\gamma^0f' \right)    - \sum_v \frac{d^2}{{d\phi_v}^2}
\end{align*}
where $m_0$ is fixed.  We then consider $H_{\cal{L} + {\cal{L}}'} + \lambda V$, with $\lambda$ small enough so that the effects of the perturbation $\lambda V$ can be analyzed within field theory.  Thus, the Dirac mass terms gap out ${\cal{L}}'$ and the trivial rotor Hamiltonian gaps out the matter degrees of freedom.

Of course the problem is that $H_{\cal{L} + {\cal{L}}'} + \lambda V$ explicitly breaks the $U(1)_A$ symmetry.  To fix this, we will couple $V$ to a dynamical order parameter $\chi$ whose quantum fluctuations will restore the $U(1)_A$ symmetry.  Specifically, we let $\chi \in C^0(M^3, \RR)$ be another set of rotors, representing the would be $U(1)_A$ order parameter.  To make $\chi$ compact, we introduce a gauge field $\beta \in C^1(M^3, \Z)$, and demand invariance under $\chi \rightarrow \chi + s, \beta \rightarrow \beta + ds$, $s\in C^1(M^3,\Z)$.  For now we view $\chi, \beta$ as classical background fields, though soon we will give them dynamics.

Consider now a $\chi$ configuration without vortices, i.e. with $d\beta=0$.  On $\RR^3$ we can represent this as the gauge equivalence class of $\beta=0$.  Consider then the unitary operator $W[\chi]$ generating a spatially non-uniform $U(1)_A$ rotation parametrized by $\chi$:
\begin{align*}
W[\chi] = \exp\left[2\pi i \int_{\RR^3} \chi \cdot \left(*j_A+*j'_A+ *j_A^{\text{matter}} \right) \right]
\end{align*}
and let
\begin{align} \label{def_of_Vchi}
V[\chi] = W[\chi]^\dagger \cdot V \cdot W[\chi].
\end{align}
Clearly $V[\chi]$ is $U(1)_V$-invariant.  For a sufficiently slowly spatially varying $\chi$ we have, explicitly
\begin{align} \label{explicitVlambda}
V[\chi] &= \int_{\RR^3}m_0\left(\cos(2\pi \chi){\psi'}^\dagger \gamma^0 \psi' + \sin(2\pi \chi){\psi'}^\dagger \gamma^0 \gamma^5 \psi'\right) \nonumber \\  &+\int_{\RR^3}m_0\left(\cos(2\pi \chi){f'}^\dagger \gamma^0 f' + \sin(2\pi \chi){f'}^\dagger \gamma^0 \gamma^5 f'\right) \nonumber\\ &- \sum_v \exp\left(-2\pi i\int_{\RR^3} \chi(x) \cdot *j_A^{\text{matter}}(x)\right)\cdot\frac{d^2}{{d\phi_v}^2} \cdot \nonumber \\ &\ \ \ \ \ \cdot \exp\left(2\pi i\int_{\RR^3} \chi(x) \cdot *j_A^{\text{matter}}(x)\right)
\end{align}
For such a slowly spatially varying configuration $\chi$, the Hamiltonian $H_{\cal{L} + {\cal{L}}'}+\lambda V[\chi]$ is gapped, because it is a sum of two decoupled pieces, each of which is individually gapped: 1) in the ancilla Hilbert space, it is just a slowly spatially varying Dirac mass term, which can be analyzed within the field theory, where it is clearly gapped, and 2) in the matter sector it is the unitary conjugate of an operator that clearly gaps the matter degrees of freedom, as seen in eq. \ref{explicitVlambda}.

Now consider a configuration $\chi$ which is still slowly varying in space, but may have some vortices.  To represent the vortices we need non-zero $\beta$; specifically, the vortex locations are along $d\beta \neq 0$.  In this case we can still construct the portion of the operator in eq. \ref{explicitVlambda} in the bulk of the system, away from the vortex locations, since there we can locally make a gauge transformation to $\beta=0$.  Let us denote this perturbation (which is $0$ at the vortex cores) by $V^{\text{bulk}}[\chi,\beta]$.  Then, with respect to the Hamiltonian $H_{\cal{L}+\cal{L}'} + \lambda V^{\text{bulk}}[\chi,\beta]$, the vortices host gapless modes in both the ancilla degrees of freedom and in the matter degrees of freedom.  This is because in both cases such vortices behave as edges of conjugate copies of $2+1$d bosonic SPTs of $U(1)_V$, arising from the fact that domain walls between different values of $\chi$ have non-zero $U(1)$ Hall conductivity proportional to the difference in $\chi$.  By design, this Hall conductivity cancels between the field theory and matter degrees of freedom, so that these gapless chiral modes can be gapped against each other with a local $U(1)_V$-invariant term $V^{\text{vortex}}[\beta]$.  Implicit in this expression is that the gapping term can be made to depend only on the vortex locations, and not on the field $\chi$.  Letting $V[\chi,\beta] = V^{\text{bulk}}[\chi,\beta] + V^{\text{vortex}}[\beta]$, we have that $H_{\cal{L} + {\cal{L}}'} + V[\chi,\beta]$ is fully gapped and $U(1)_V$-invariant for all configurations $\chi, \beta$.

For a spatially constant $\chi_0$, $W[\chi_0]$ just generates a global $U(1)_A$ rotation by $\chi_0$.  From eq. \ref{def_of_Vchi} and the definition of $V[\chi,\beta]$, we have $W[\chi_0]^\dagger \, V[\chi,\beta] \, W[\chi_0] = V[\chi+\chi_0,\beta]$.  This means that if we now make $\chi$ and $\beta$ dynamical, and assign a $U(1)_A$ charge of $-1$ to $\chi$, then the Hamiltonian $H_{\cal{L} + {\cal{L}}'} + V[\chi,\beta]$ is $U(1)_A$ invariant.  Formally, to make $\chi,\beta$ dynamical we have to introduce another Hilbert space ${\cal{H}_{\chi}}$, so that the total Hilbert space is ${\cal{H}_{\chi}} \otimes {\cal{H}_{\cal{L} + \cal{L}'}} \otimes {\cal{H}}_{\text{matter}}$, with ${\cal{H}_{\chi}} \otimes {\cal{H}_{\cal{L} + \cal{L}'}}$ acting as ancillas.

Even though the Hamiltonian $H_{\cal{L} + {\cal{L}}'} + V[\chi,\beta]$ is now $U(1)_A$ invariant, $\chi$ and $\beta$ are still background fields with no dynamics.  In order to fully restore $U(1)_A$ we have to make these fields fluctuate, and ensure that the resulting Hamiltonian does not spontaneously break the $U(1)_A$.  Consider
\begin{align} \label{Ham_ungauged}
H &= H_{\cal{L} + {\cal{L}}'}+\lambda V[\chi,\beta] + H_{\chi,\beta}
\end{align}
where $H_{\chi,\beta}$ will be a term that acts only on the $\chi,\beta$ degrees of freedom.  To fix the form of $H_{\chi,\beta}$, we first consider just the two terms $H_{\cal{L} + {\cal{L}}'}+\lambda V[\chi,\beta]$, and integrate out the degrees of freedom in ${\cal{H}}_{\cal{L}+\cal{L}'}$ and ${\cal{H}}^{\text{matter}}$ to get an effective potential for $\chi,\beta$.  Even though there are still gapless ${\cal{L}}$ degrees of freedom, they do not couple to $\chi$ and $\beta$, so this effective potential is short-ranged.  We then choose $H_{\chi,\beta}$ to exactly cancel out this effective potential, and also include a small kinetic term for $\chi$ and $\beta$.  The effective long distance description of the $\chi,\beta$ degrees of freedom is then as an $XY$ model in the quantum-disordered phase.  This quantum-disordering restores the $U(1)_A$ symmetry and gives us our desired realization of a single copy of ${\cal{L}}$, in the Hilbert space ${\cal{H}_{\chi}} \otimes {\cal{H}_{\cal{L} + \cal{L}'}} \otimes {\cal{H}}_{\text{matter}}$.

\subsection{Gauged $U(1)_V$}

Let us now gauge the $U(1)_V$ symmetry by weakly coupling the Lagrangian $L$ in eq. \ref{SU2Lagrangian} to a dynamical $U(1)_V$ gauge field.  In this field theory the $U(1)_A$ is no longer a symmetry, but a $\Z_N$ remnant remains as a non-invertible symmetry, as is the case in QED and other standard examples of non-invertible symmetries \cite{Shao_TASI}.  We will work on $M^3=\RR^3$.  Since $\RR^3$ has no non-trivial $2$-cycles, and the field theory contains no monopoles, the action of the non-invertible symmetry on this particular Hilbert space is actually unitary.

We argue that this theory can emerge from an appropriate Hamiltonian on a slightly modified version of our Villain Hilbert space $\Hg$, which results from gauging $U(1)_V$ not just on ${\cal{H}}_{\text{matter}}$, but on ${\cal{H}_{\chi}} \otimes {\cal{H}_{\cal{L} + \cal{L}'}} \otimes {\cal{H}}_{\text{matter}}$.  Since both $U(1)_V$ and $U(1)_A$ act in an onsite fashion in the ancilla Hilbert space ${\cal{H}_{\chi}} \otimes {\cal{H}_{\cal{L} + \cal{L}'}}$, this is just a trivial modification of the Villain Hilbert space obtained from gauging $U(1)_V$ in just ${\cal{H}}_{\text{matter}}$.  The end result is a Hilbert space very similar to that of our Villain Hilbert space $\Hg$, but with additional ancilla superspin degrees of freedom living on vertices and transforming in some well-defined onsite representations of $U(1)_V$ and $U(1)_A$.

The first step in our construction is to take zero coupling, i.e. no dynamics for the gauge field $a$, so that we can view $a$ as a static spatial background gauge field. 
Also, we will include an energetic penalty for monopole configurations, and the non-trivial terms in the Hamiltonian will all be sandwiched between projectors onto zero monopoles, and commute with such projectors.  We then gauge the Hamiltonian $H$ in eq. \ref{Ham_ungauged} by following the minimal coupling prescription, resulting in $H^{\text{gauged}}$.  The $U(1)_A$ symmetry of the un-gauged theory turns into an $\RR$ symmetry for this gauged Hamiltonian.  Note that we are still working at zero coupling, i.e. with no fluctuations for $a$; a non-zero electric field term would fail to commute with this $\RR$ symmetry, a reflection of the ABJ anomaly.

$H^{\text{gauged}}$ is also invariant under the non-invertible symmetry $U_{\frac{1}{N}}(\RR^3)$.  This is because the difference between $U_{\frac{1}{N}}(\RR^3)$ and an ordinary $U(1)_A$ rotation by $\frac{2\pi}{N}$, $\exp(-\frac{2\pi i}{N} Q_5) U_{\frac{1}{N}}(\RR^3)$, is entirely independent of the charged (matter) fields, i.e. depends only on the gauge field configuration.  Since $H^{\text{gauged}}$ does not contain any electric field terms, it commutes with this operator, and hence with $U_{\frac{1}{N}}(\RR^3)$.

However, if we are only concerned with retaining invariance under $U_{\frac{1}{N}}(\RR^3)$, we can now introduce some fluctuations of $a$.  We will use the projected electric field operator $P_{\Hgm}\cdot \frac{d}{da_e} \cdot P_{\Hgm}$, which induces fluctuations of the vector potential within the subspace $\Hgm$.  This term does not commute with $U_{\frac{1}{N}}(\RR^3)$, but if we conjugate it by $U_{\frac{1}{N}}(\RR^3)^k$ and sum over $k = 0,1,\ldots,N-1$, we get a term that trivially commutes with $U_{\frac{1}{N}}(\RR^3)$.  Since the eigenvalues of $U_{\frac{1}{N}}(\RR^3)$ are discrete, this electric field term delocalizes the vector potential $a$ over the configuration space of possible $a$'s corresponding to a particular eigenvalue of $U_{\frac{1}{N}}(\RR^3)$ (which also depends on $\phi$).  If we also add a magnetic field term, then at low energies this simply flows to the theory ${\cal{L}}$, weakly coupled to a $U(1)_V$ gauge field, as desired.

\section{Conclusions} \label{sec:conclusions}

In this work we first considered a $3+1$ d tensor product Hilbert space of rotors, with an onsite action of a $U(1)_V$ symmetry that rotates the rotors.  We constructed a non-onsite constant depth circuit action of an axial $U(1)_A$ symmetry on this Hilbert space, which had the form of $\exp\left[i\theta \int_{M^3} *j_A^{\text{matter}}\right]$, where $*j_A^{\text{matter}}$ is a collection of mutually commuting operators that define an axial charge density.  The defining property of this $U(1)_A$ symmetry was a mixed anomaly with $U(1)_V$, which manifests itself on the lattice through the fact that performing an axial rotation in a region pumps Hall conductivity out to the boundary of the region.

We then gauged this $U(1)_V$ symmetry at the level of the Hamiltonian lattice, and obtained a gauged axial current, $*j_A$.  This gauged axial current did not generate a $U(1)$ symmetry, but rather an $\RR$ symmetry: $\exp\left[i \theta \int_{M^3} *j_A\right]$, $\theta\in \RR$.  Also, it was not conserved, because it failed to commute with the electric field operator.  However, focusing on just the subspace $\Hgm \subset \Hg$ of states invariant under a $\Z_N$ one-form magnetic symmetry $Z_m^{(1)}$, we were able to find a $\Z_N$ group of locality preserving unitary operators whose action on the charged matter degrees of freedom matched that of $\exp\left[\frac{2\pi ik}{N}\theta \int_{M^3} *j_A\right]$, $k \in \{0,1,\ldots, N-1\}$.  The generator of this group $U_{\frac{1}{N}}(M^3)$ involved a projection onto $\Hgm$, and was our definition of the non-invertible chiral symmetry.

We were able to shed more light on the nature of $U_{\frac{1}{N}}(M^3)$ by gauging $Z_m^{(1)}$ to obtain a dual Hilbert space $\Hge \cong \Hgm$.  $\Hge$ was the subspace of $\HgN$, the Hilbert space of a charge $\frac{1}{N}$ scalar coupled to a $U(1)_V$ gauge field.  Specifically, it was the subspace invariant under a $\Z_N$ electric one-form symmetry $Z_e^{(1)}$. 
 $U_{\frac{1}{N}}(M^3)$, viewed as an operator on $\Hge$, extended to a constant depth circuit of local unitaries on $\HgN$ with a mixed anomaly with $Z_e^{(1)}$; in particular, it entangled an SPT of $Z_e^{(1)}$.

There are several avenues for further investigation.  One is to address gravitational anomalies within this framework.  For example, one may ask whether there is a circuit which pumps out a $2+1$d phase with non-zero thermal Hall response, if one allows for infinite dimensional site Hilbert spaces.  This hinges on whether it is possible to construct a commuting projector Hamiltonian for a $2+1$d state with non-zero thermal Hall conductivity, again with infinite dimensional sites.  Perhaps an intermediate step is asking if there is a circuit that pumps out 2+1d phases with zero thermal Hall response but no gappable boundary \cite{levin2013}.  Additionally, within the context of the present work, one can attempt to understand the full $SL(2,\Z_N)$ structure \cite{Choi_standard_model} that should be generated by the one-form gauging operator $\rho$ (which plays the role of `$S$') and $U_{\frac{1}{N}}(M^3)$ acting in $\Hge$ (which plays the role of `$T$').  In particular, while $U_{\frac{1}{N}}(M^3)$ acts as the Tsui-Wen SPT entangler in $\Hge$ \cite{Tsui}, its action in $\Hgm$, which should be viewed as the conjugate of $T$ by $S$, is a Walker-Wang entangler for the chiral $\Z_N$ anyon theory.  This means that it also shifts the gravitational theta angle, as can be seen from the non-zero chiral central charge of the chiral $\Z_N$ anyon theory.  It would be interesting to relate this to the projective action of $SL(2,\Z_N)$ discussed in ref. \cite{Choi_standard_model}.  Additionally, it would be interesting to relate this structure to the $\Z_N$ QCA found in ref. \cite{YuAnQCA2025}, which are {\emph{invertible}} operators that produce the Walker-Wang ground state. There has also been related work on non-invertible time-reversal symmetry\cite{choi2023T}; the lattice implementation of such operators appears similar to the non-invertible axial rotations described in this work.

Another direction to explore is a fermionic version of this story.  This is natural since the simplest formulation of the ABJ anomaly involves a $U(1)_V$ gauge theory with fermionic charges, namely QED. The corresponding `un-gauged' theory is just a Dirac fermion, as opposed to our more complicated bosonic theory in eq. \ref{SU2Lagrangian_first}.  In order to cancel the gravitational contribution to the $U(1)_V$ anomaly, we could consider a charged Dirac fermion stacked together with a neutral one, with opposite actions of $U(1)_A$.  This has the advantage of being a free fermion theory, and it is therefore trivial to realize this theory stacked with its conjugate, with the diagonal $U(1)_V$ and $U(1)_A$ symmetries acting onsite, in a microscopic lattice Hamiltonian.  The existence of such a realization in the bosonic case was a crucial assumption in our emergeability argument in section \ref{sec:FieldTheory}; in the fermionic case we do not need to make such an assumption.  To build the microscopic fermionic Hilbert space with an action of $U(1)_A$, we need to introduce lattice fermion degrees of freedom, and write down a generating time independent charge that pumps out an integer quantum Hall state of charged fermions stacked with an opposite Chern number integer quantum Hall state of neutral fermions.

Another avenue of investigation that is broadly motivated by this work is to compare the classification of QCA in Hilbert spaces built out finite dimensional sites versus infinite dimensional rotors.  The operator $*j_A^{\text{matter}}$ constructed in this work can be thought of as a generator of a loop in in the space of $3$d QCA, and hence gives rise to a $2$d $U(1)$-protected QCA at the boundary.  This $2$d $U(1)$-protected QCA just entangles the bosonic $U(1)$-protected SPT state, and is only allowed to exist in this rotor setting, since it is known that with finite dimensional site Hilbert spaces, it is not possible to have a commuting projector parent Hamiltonian for a system with non-zero $U(1)$ Hall conductivity \cite{KapustinFidkowski2019, ZhangLevin2022}. 

It would also be interesting to make a more direct connection between our lattice methods and the corresponding structures in field theory.  For example, in section \ref{sec:U1A} we constructed the operator $*j_A = *j_A^{(I)} + *j_A^{(II)}$.  $*j_A$ was a well-defined operator in $\Hg$, but the individual pieces $*j_A^{(I)}$ and $*j_A^{(II)}$ were not.  Nevertheless, $\exp\left[\frac{2\pi i}{N}\int_{M^3} *j_A^{(I)} \right]$ and $\exp\left[\frac{2\pi i}{N}\int_{M^3} *j_A^{(II)} \right]$ were individually well defined in $\Hge \subset \HgN$, for closed $M^3$.  The latter was just our non-invertible symmetry operator $U_{\frac{1}{N}}(M^3)$, which performs a chiral rotation on the matter fields.  The former does not depend on the matter field $\phi$ (for gauge field configurations with no monopoles), and can be viewed as performing a shift of the theta angle.  Both of these are non-trivial QCA in $\Hge$ (and therefore also $\Hgm$), whereas their product is a trivial QCA, since $*j_A$ are well-defined local operators.  This comports with the field theory idea that a chiral rotation can be `un-done' by a shift of theta angle; from our lattice perspective, both of these operations are non-trivial QCA, but taken together they form a trivial QCA, i.e. a circuit.

\vspace{0.5cm}

{\bf{Acknowledgements:}} We thank the KITP for hospitality during the program GenSym25, during which this collaboration was initiated. L.F. is supported by NSF DMR-
2300172.  C.Z. is supported by the Harvard Society of Fellows. C.X. is supported by the Simons Foundation through the Simons Investigator program.

\appendix

\section{Details on gauging $\Z_N$ one-form symmetries} \label{app:gauging_appendix}
\subsection{Definition of inner product in $\Hg$}

Due to the Villain condition (eq. \ref{Villain}) wavefunctions $\Psi(\phi,a,n)$ in $\Hg$ appear to be non-normalizable if one uses the naive norm, obtained from integrating $|\Psi(\phi,a,n)|^2$ over all $\phi,a$ and summing over $n$.  However, the Villain condition should clearly just be treated as a redundancy when defining the inner product.  One way to handle this is to use the Villain condition to move the variables $(a-d\phi)_e$ into the range $[0,1)$ for all edges $e$.  Each orbit under the Villain condition has precisely one gauge choice where all of the $(a-d\phi)_e$ are in this range.  This results in an inner product
\begin{align} \label{innerproduct}
\langle \Psi|\Psi'\rangle = \int_0^1 {\cal{D}}(a-d\phi)\sum_n{\overline{\Psi(\phi,a,n)}}\cdot \Psi'(\phi,a,n)
\end{align}
where
\begin{align*}
\int_0^1 {\cal{D}}(a-d\phi) \equiv \prod_{\text{edges } e}\int_0^1 d(a-d\phi)_e
\end{align*}
It will also be useful to have an expression for this product in a Fourier-transformed `plane-wave' basis on the configuration space $\phi,a,n$.  Let $\ve \in C^1(M^3,\RR), \vt \in C^2(M^3,\RR/\Z)$, and consider the state $|e,\vt\rangle$, defined by the wavefunction
\begin{align*}
\exp \left[2\pi i\int_{M^3}\left(-\partial \ve \cdot \phi+\ve\cdot a + \vt\cdot n \right) \right]
\end{align*}
where we use the notation $a \cdot b \equiv a \cup *b$.  The Villain condition in eq. \ref{Villain} then turns into the condition $\ve + \partial \vt \in C^1(M^3,\Z)$, where $\partial = *\,d\,*$ and $*$ is the Hodge dual.  To understand this, note that integer translations of $a$ and $n$, applied to the plane wave state above, yield phase factors of $\exp(2\pi i \ve)$ and $\exp(2\pi i\vt)$ respectively.  Then the Villain condition in eq. \ref{Villain} turns into the requirement that for all $v \in C^1(M^3, \Z)$
\begin{align} \label{dual_constraint}
\int_{M^3} \left( v\cdot \ve + dv \cdot \vt\right) \in \Z
\end{align}
We have that $\int dv \cdot \vt = \int v \cdot \partial \vt$, so eq. \ref{dual_constraint} turns into the requirement that $\ve + \partial \vt \in C^1(M^3,\Z)$.  This is just the statement of charge quantization in the Villain formalism.

A general state in our Hilbert space can be written as
\begin{align*}
|\Psi\rangle = \int_0^1{\cal{D}}\vt \sum_{\substack{\ve \in C^1(M^3,\RR)\,|\\ \ve+\partial\vt\in C^1(M^3,\Z)}} \Psi_{\ve,\vt} \,|\ve,\vt\rangle
\end{align*}
where $\int_0^1{\cal{D}}\vt$ is shorthand for $\prod_f\int_0^1 d\vt_f$ and the product is over all faces.  The sum on $\ve$ is due to the constraint that $\ve+\partial \vt$ be integer valued; for a given configuration $\vt$, the allowed values of $\ve$ are quantized up to $-\partial \vt$.  We could have written the above as an integral over both $\vt$ and $\ve$, but then we would have had to include delta functions imposing these constraints inside the integral.  The wavefunction of the state $|\Psi\rangle$ is
\begin{align} \label{wf1}
\Psi(\phi,a,n)=\int_0^1{\cal{D}}\vt\sum_{\substack{\ve \in C^1(M^3,\RR)\,|\\ \ve+\partial\vt\in C^1(M^3,\Z)}}\Psi_{\ve,\vt} \cdot\exp \left[2\pi i\int_{M^3}\left(-\partial \ve \cdot \phi+\ve\cdot a + \vt\cdot n \right) \right]
\end{align}
Conversely, given $\Psi(\phi,a,n)$ invariant under the usual gauge invariance and the Villain condition, we can recover $\Psi_{\ve,\vt}$ by taking the inverse Fourier transform:
\begin{align*}
\Psi_{\ve,\vt} = \int_{0}^1 {\cal{D}}(a-d\phi) \sum_{n\in C^2(M^3,\Z)} \exp\left[-2\pi i \int_{M^3} \left(\ve\cdot(a-d\phi) + \vt\cdot n\right)\right] \Psi(\phi,a,n)
\end{align*}
Here $\int_{0}^1 {\cal{D}}(a-d\phi)$ is shorthand for $\prod_e \int_0^1 d(a-d\phi)_e$ where the product is over all edges $e$.

From taking the mod squared of the above equation and Fourier-transforming, we obtain that
\begin{align*}
\int_0^1 {\cal{D}}\vt \sum_{\substack{\ve \in C^1(M^3,\RR)\,|\\ \ve+\partial\vt\in C^1(M^3,\Z)}} |\Psi_{\ve,\vt}|^2 = \int_0^1 {\cal{D}}(a-d\phi)\sum_n|\Psi(\phi,a,n)|^2
\end{align*}
Therefore an alternative formula for the inner product in eq. \ref{innerproduct} is
\begin{align*}
\langle \Psi |\Psi'\rangle \equiv \int_0^1 {\cal{D}}\vt \sum_{\substack{\ve \in C^1(M^3,\RR)\,|\\ \ve+\partial\vt\in C^1(M^3,\Z)}} {\overline{\Psi_{\ve,\vt}}}\, \Psi'_{\ve,\vt}
\end{align*}

Let us also discuss the Hilbert space $\HgN$ of the gauge theory of a charge $\frac{1}{N}$ scalar.  Here $n$ is $N\Z$-valued, and the invariance is only under the restricted Villain transformation (eq. \ref{mod_Villain}).  By the same argument as for ${\cal{H}}$, we can take account of this restricted Villain transformation and define the inner product as:
\begin{align} \label{Ninnerproduct}
\langle \Psi^N|{\Psi'}^N\rangle_N = \int_0^N {\cal{D}}(a-d\phi)\sum_{n\in C^2(M^3,N\Z)}{\overline{\Psi^N(\phi,a,n)}}\cdot {\Psi'}^N(\phi,a,n)
\end{align}

The corresponding dual representation has $\ve \in C^1(M^3,\RR), \vt \in C^2(M^3, \RR / (\frac{1}{N}\Z))$, and the modified Villain condition (eq. \ref{mod_Villain}) turns into $\ve+\partial \vt \in C^1(M^3,\frac{1}{N}\Z)$.  A general state in this Hilbert space is
\begin{align*}
|\Psi^N\rangle = \int_0^{\frac{1}{N}}{\cal{D}}\vt \sum_{\substack{\ve \in C^1(M^3,\RR)\,|\\ \ve+\partial\vt\in C^1(M^3,\frac{1}{N}\Z)}} \Psi^N_{\ve,\vt} \,|\ve,\vt\rangle_N
\end{align*}
where $\int_0^{\frac{1}{N}}{\cal{D}}\vt$ is shorthand for $\prod_f\int_0^{\frac{1}{N}} d\vt_f$.  Here the state $|\ve,\vt\rangle_N$ is given by the wavefunction
\begin{align*}
\exp \left[2\pi i\int_{M^3}\left(-\partial \ve \cdot \phi+\ve\cdot a + \vt\cdot n \right) \right]
\end{align*}
with $n \in C^2(M^3,N\Z)$.
The wavefunction of the state $|\Psi^N\rangle$ is
\begin{align} \label{wfN}
\Psi^N(\phi,a,n)=\int_0^{\frac{1}{N}}{\cal{D}}\vt\sum_{\substack{\ve \in C^1(M^3,\RR)\,|\\ \ve+\partial\vt\in C^1(M^3,\frac{1}{N}\Z)}}\Psi_{\ve,\vt}^N \cdot\exp \left[2\pi i\int_{M^3}\left(-\partial \ve \cdot \phi+\ve\cdot a + \vt\cdot n \right) \right]
\end{align}
where again $n \in C^2(M^3,N\Z)$.  Conversely, given $\Psi^N(\phi,a,n)$ invariant under the usual gauge invariance and the modified Villain condition in eq. \ref{mod_Villain}, we can recover $\Psi^N_{\ve,\vt}$ by
\begin{align*}
\Psi^N_{\ve,\vt} = N^{|\text{faces}|-|\text{edges}|} &\int_{0}^N {\cal{D}}(a-d\phi) \sum_{n\in C^2(M^3,N\Z)} \\ &\exp\left[-2\pi i \int_{M^3} \left(\ve\cdot(a-d\phi) + \vt\cdot n\right)\right]\,\Psi^N(\phi,a,n)
\end{align*}
where the extra factor out front comes from changing the quantization for the fields $n,a$.  Here $\int_{0}^N {\cal{D}}(a-d\phi)$ is shorthand for $\prod_e \int_0^N d(a-d\phi)_e$ where the product is over all edges $e$.

By taking the mod squared of both sides of the above equation and Fourier transforming, we obtain:
\begin{align*}
\int_0^{\frac{1}{N}}& {\cal{D}}\vt \sum_{\substack{\ve \in C^1(M^3,\RR)\,|\\ \ve+\partial\vt\in C^1(M^3,\frac{1}{N}\Z)}} |\Psi^N_{\ve,\vt}|^2 \\ &= N^{|\text{faces}|-|\text{edges}|} \int_0^N {\cal{D}}(a-d\phi)\sum_{n\in C^2(M^3, N\Z)}|\Psi(\phi,a,n)|^2
\end{align*}
Therefore the inner product in eq. \ref{Ninnerproduct} can alternatively be written as
\begin{align} 
\langle \Psi^N|{\Psi'}^N\rangle_N \equiv N^{|\text{edges}|-|\text{faces}|}\int_0^{\frac{1}{N}} {\cal{D}}\vt \sum_{\substack{\ve \in C^1(M^3,\RR)\,|\\ \ve+\partial\vt\in C^1(M^3,\frac{1}{N}\Z)}} {\overline{\Psi_{\ve,\vt}^N}}\,{\Psi'}^N_{\ve,\vt}
\end{align}

\subsection{Structure of $\Hge \subset \HgN$}

Let us now take a state $|\Psi\rangle \in \Hge \subset \HgN$:
\begin{align*}
|\Psi^N\rangle = \int_0^{\frac{1}{N}}{\cal{D}}\vt \sum_{\substack{\ve \in C^1(M^3,\RR)\,|\\ \ve+\partial\vt\in C^1(M^3,\frac{1}{N}\Z)}} \Psi^N_{e,\vt} \,|\ve,\vt\rangle_N, 
\end{align*}
We already know from the rescaled Villain constraint (eq. \ref{mod_Villain}) that $\ve+\partial \vt \in C^1(M^3, \frac{1}{N}\Z)$.  From the definition of $Z_e^{(1)}$ in eq. \ref{ZNe} we find that invariance under $Z_e^{(1)}$ places additional constraints on which of the coefficients $\Psi_{\ve,\vt}$ can be non-zero: namely, $\int_{M^3} v \cdot (\ve+\partial\vt)$ must be an integer for any $v \in C^1(M^3,\Z)$ satisfying $dv \in C^2(M^3, N\Z)$.  Taking $v=ds$ for $s \in C^0(M^3,\Z)$ gives $\int_{M^3} s \cdot \partial(N(\ve+\partial\vt)) \in N\Z$ for all such $s$, so that $\partial(N(\ve+\partial\vt)) \in C^0(M^3,N\Z)$.  This means that $*N(\ve+\partial\vt)$ is a $\Z_N$-valued 2-cocycle, where $\Z_N = \{0,1,\ldots, N-1\}$.  Furthermore, given any closed $\Z_N$ 2-chain $\Sigma \in M^3$, let $v = \text{PD}(\Sigma) \in C^1(M^3,\Z_N)$ be its Poincare dual.  Then $dv = 0$ mod $N$, so:
\begin{align*}
\int_\Sigma *(N(\ve+\partial\vt)) = N\int_{M^3}v\cdot(\ve+\partial\vt) \in N\Z
\end{align*}
Since this is true for all closed $\Sigma$, $*N(\ve+\partial\vt)$ is exact, implying that $N(\ve+\partial\vt) = \partial \beta$ for some $\beta \in C^2(M^3,\Z_N)$.  Thus, while the constraint on a general state in $\HgN$ is that $\ve+\partial\vt$ be valued in $\frac{1}{N} \Z$, for a state in $\Hge \subset \HgN$ we have the additional constraint that $\ve + \partial\vt = \frac{1}{N} \partial \beta$ mod $1$, for some integer-valued $\beta$.  We will see below that $\partial \beta$ can be interpreted as a pure-gauge dual $2$-form gauge field obtained from gauging $\Z_m^{(1)}$ symmetry in $\Hgm$.

\subsection{Gauging $\Z_m^{(1)}$ by minimal coupling} \label{subsec:gauging}

The action of the $Z_m^{(1)}$ symmetry in the Fourier transformed basis for $\Hg$ is:
\begin{align} \label{dualZmaction}
Z_m^{(1)}(\Sigma)|\ve,\vt\rangle = |\ve,\vt + \frac{1}{N}\cdot*\text{PD}(\Sigma)\rangle
\end{align}
where $\text{PD}(\Sigma)$ is the Poincare dual of $\Sigma$.  Let us now gauge $Z^{(1)}_m$ by coupling to a pure gauge $\Z_N$ $2$-form gauge field.  Since the $Z^{(1)}_m$ 1-form symmetry acts on the dual edges, its gauge field will live on dual plaquettes, i.e. ordinary edges.  Its being pure gauge means that it is equal to $\partial\beta =\!\ast d\!\ast\beta$, where $\beta \in C^2(M^3,\Z_N)$; in the formulae below we treat $\beta$ as being valued in the integers $0,\ldots, N-1$.  The wavefunctions of the $Z^{(1)}_m$-gauged theory now include dependence on the $Z^{(1)}_m$ gauge field $\partial \beta$:
\begin{align} \label{gaugedWf}
|\Psi_g\rangle=\sum_{\partial\beta \in C^1(M^3,\Z_N) }\int_0^{1}{\cal{D}}\vt \sum_{\substack{\ve \in C^1(M^3,\RR)\,|\\ \ve+\partial\vt -\frac{1}{N}\partial \beta \in C^1(M^3,\Z)}} \Psi_{\ve,\vt,\partial\beta} \,|\ve,\vt, \partial\beta\rangle
\end{align}
Let us explain this formula.  $|\Psi_g\rangle$ is a state in the Hilbert space of $\Hg$ tensored with that of a $\Z_N$ degree of freedom living in $C^1(M^3,\Z_N)$.  The first sum above is over all dual-exact such $1$-cochains, i.e. all such cochains of the form $\partial\beta$ for some $\beta$.  The inner product on this Hilbert space is inherited from ${\cal{H}}$ and the natural inner product on the $\Z_N$ degrees of freedom.  We will explain the sum on $\ve$ momentarily.

$|\Psi_g\rangle$ must be invariant under the $Z^{(1)}_m$-Gauss law:
\begin{align} \label{Zm1_Gauss}
\Psi_{\ve,\vt+\frac{1}{N}\cdot\beta', \partial \beta + \partial \beta'} = \Psi_{\ve,\vt,\partial\beta}
\end{align}
for any co-chain $\beta' \in C^2(M^3, \Z_N)$.  The Villain constraint in eq. \ref{dual_constraint} also needs to be conjugated by the $Z^{(1)}_m$-Gauss law, leading to the sum over $\ve$ in eq. \ref{gaugedWf} being restricted to
\begin{align} \label{gauged_dual_constraint}
\ve + \partial\vt -\frac{1}{N}\partial \beta \in C^1(M^3,\Z)
\end{align}
Note that this means that $\partial \beta$ is completely determined, as a $\Z_N$ valued cochain, by the configuration of $\ve, \vt$.

We now see that the following map is a unitary isomorphism between the space of pure-gauge wavefunctions in the $Z_m^{(1)}$-gauged theory and $\Hge$:
\begin{align} \label{forwardmap}
\sum_{\partial\beta \in C^1(M^3,\Z_N) }&\int_0^{1}{\cal{D}}\vt \sum_{\substack{\ve \in C^1(M^3,\RR)\,|\\ \ve+\partial\vt -\frac{1}{N}\partial \beta \in C^1(M^3,\Z)}} \Psi_{\ve,\vt,\partial\beta} \,|\ve,\vt, \partial\beta\rangle \rightarrow \nonumber \\ & N^{(|\text{faces}|-|\text{edges}|)/2}\int_0^{\frac{1}{N}}{\cal{D}}\vt \sum_{\substack{\ve \in C^1(M^3,\RR)\,|\\ \ve+\partial\vt = \frac{1}{N}\partial\beta \text{ mod }1 \\ \text{for some } \beta\in C^2(M^3,\Z_N)}} \Psi_{\ve,\vt, N(e+\partial\vt)} \,|\ve,\vt\rangle_N
\end{align}
with inverse given by:
\begin{align*}
\int_0^{\frac{1}{N}}{\cal{D}}\vt &\sum_{\substack{\ve \in C^1(M^3,\RR)\,|\\ \ve+\partial\vt = \frac{1}{N}\partial\beta \text{ mod }1 \\ \text{for some } \beta\in C^2(M^3,\Z_N)}} \Psi^N_{\ve,\vt} \,|\ve,\vt\rangle_N \rightarrow \\ &N^{(-|\text{faces}|+|\text{edges}|)/2}  \int_0^{1}{\cal{D}}\vt \sum_{\substack{\ve \in C^1(M^3,\RR)\,|\\ \ve+\partial\vt = \frac{1}{N}\partial\beta \text{ mod }1 \\ \text{for some } \beta\in C^2(M^3,\Z_N)}} \Psi^N_{\ve,\vt} \,|\ve,\vt, N(\ve+\partial\vt)\rangle
\end{align*}
Note that the wavefunction on the right hand side of eq. \ref{forwardmap} is invariant under $\frac{1}{N}$ shifts of $\vt$, due to the $\Z_m^{(1)}$ Gauss law (eq. \ref{Zm1_Gauss}).

\subsection{Isomorphism $\Hgm \rightarrow \Hge$ from gauging $\Z_m^{(1)}$} \label{subsec:HgmHge}
Now, given any state
\begin{align*}
 \int_0^1{\cal{D}}\vt \sum_{\substack{\ve \in C^1(M^3,\RR)\,|\\ \ve+\partial\vt \in C^1(M^3,\Z)}} \Psi_{\ve,\vt} \,|\ve,\vt\rangle \in \Hgm,
\end{align*}
we can view it as a state coupled to the trivial background gauge field configuration $\partial\beta=0$.  We can then make it gauge invariant with respect to the $\Z_m^{(1)}$ Gauss law (eq. \ref{Zm1_Gauss}) by allowing the gauge field to fluctuate over all pure gauge configurations.  We accomplish this by mapping it to the state:
\begin{align*}
\frac{1}{\sqrt{|{\{\partial\beta\}|}}}\sum_{\partial\beta}\int_0^1{\cal{D}}\vt \sum_{\substack{\ve \in C^1(M^3,\RR)\,|\\ \ve+\partial\vt -\frac{1}{N}\partial \beta \in C^1(M^3,\Z)}} \Psi_{\ve,\vt - \frac{1}{N}\beta} \,|\ve,\vt, \partial \beta\rangle.
\end{align*}
This map is unitary.  Note that $\Psi_{\ve,\vt - \frac{1}{N}\beta}$ depends only on $\partial \beta$ by virtue of $Z_m^{(1)}$-invariance (eq. \ref{dualZmaction}), so the sum above is really over exact cochains $d\beta$.  Composing with the map in eq. \ref{forwardmap}, we get a unitary isomorphism, denoted $\rho$ in the main text, from $\Hgm$ to $\Hge$:
\begin{align*}
 \int_0^1{\cal{D}}&\vt \sum_{\substack{\ve \in C^1(M^3,\RR)\,|\\ \ve+\partial\vt \in C^1(M^3,\Z)}} \Psi_{\ve,\vt} \,|\ve,\vt\rangle \rightarrow \\ &  \frac{N^{(|\text{faces}|-|\text{edges}|)/2}}{\sqrt{|{\{\partial\beta\}|}}} \int_0^{\frac{1}{N}}{\cal{D}}\vt \sum_{\substack{\ve \in C^1(M^3,\RR)\,|\\ \ve+\partial\vt = \frac{1}{N}\partial\beta \text{ mod }1 \\ \text{for some } \beta\in C^2(M^3,\Z_N)}} \Psi_{\ve,\vt-\frac{1}{N}\beta} \,|\ve,\vt\rangle_N.
\end{align*}
The map $\rho$ is especially simple in the original $(\phi,a,n)$ basis.  Applying eq. \ref{wfN} to the right hand side of the above (but without yet including the overall constant $\frac{N^{|\text{faces}|-|\text{edges}|/2}}{\sqrt{|{\{\partial\beta\}|}}}$), we have, for $n \in C^2(M^3, N\Z)$,
\begin{align*}
&\int_0^{\frac{1}{N}}{\cal{D}}\vt \sum_{\substack{\ve \in C^1(M^3,\RR)\,|\\ \ve+\partial\vt = \frac{1}{N}\partial\beta \text{ mod }1 \\ \text{for some } \beta\in C^2(M^3,\Z_N)}} \Psi_{\ve,\vt-\frac{1}{N}\beta} \exp\left(2\pi i \int_{M^3}\left(\ve\cdot(a-d\phi)+\vt \cdot n\right)\right)= \\
&\frac{1}{|\{\beta|\partial\beta=0\}|}\int_0^{1}{\cal{D}}\tilde\vt \sum_{\substack{\ve \in C^1(M^3,\RR)\,|\\ \ve+\partial\tilde\vt \in C^1(M^3,\Z)}} \psi_{\ve,\tilde\vt}\exp\left(2\pi i \int_{M^3}\left(\ve\cdot(a-d\phi)+\vt \cdot n\right)\right)= \\
&\frac{1}{|\{\beta|\partial\beta=0\}|} \Psi(\phi,a,n)
\end{align*}
Thus
\begin{align}\label{derivedrho}
\rho(\Psi)(\phi,a,n)&=\frac{N^{(|\text{faces}|-|\text{edges}|)/2}}{|\{\beta|\partial\beta=0\}|\cdot\sqrt{|{\{\partial\beta\}|}}} \Psi(\phi,a,n) \\ &= \frac{N^{(|\text{faces}|-|\text{edges}|)/2}}{\sqrt{|{\{\beta|\partial\beta=0\}|}}} \Psi(\phi,a,n)
\end{align}
This is equation \ref{def_rho} in the main text.  One can explicitly check that the map in eq. \ref{derivedrho} is unitary by using the original definitions of the inner product in the $(\phi,a,n)$ basis (eqs. \ref{innerproduct} and \ref{Ninnerproduct}).  This check involves showing that a combination of the above prefactor and other constants is equal to $1$ using the fact that the Euler characteristic of an odd dimensional manifold vanishes, and Hodge duality, which relates the first cohomology and second homology groups of $M^3$ with $\Z_N$ coefficients.

\subsection{Isomorphism $\Hge \rightarrow \Hgm$ from gauging $\Z_e^{(1)}$} \label{subsec:HgeHgm}
For completeness, we also show that gauging $Z^{(1)}_e$ (with a pure gauge gauge field) recovers $\Hgm$, and that the resulting isomorphism $\Hge \rightarrow \Hgm$ is inverse to that constructed in eq. \ref{def_rho}.  Gauging $Z^{(1)}_e$ requires the introduction of $\Z_N$-valued $2$-form degrees of freedom in $\HgN$.  We do this simply by extending $n$ from being $N\Z$-valued to being $\Z$-valued.  Then $n \text{ mod } N$ is the $\Z_N$ $2$-form gauge field for $Z_e^{(1)}$.  The second step is to impose a $Z^{(1)}_e$-Gauss law on each edge $e$, which just amounts to the ordinary Villain condition (eq. \ref{Villain}).  The explicit map at the level of wave-functions is as follows. Given $\Psi(\phi,a,n) \in \Hge$, we view it as a wave-function in which $n$ is allowed to take on all integer values (and we set its value equal to $0$ for any configuration involving $n\neq 0$ mod $N$).  Then we symmetrize it with respect to the $Z^{(1)}_e$-Gauss law.  This symmetrization amounts to the following: given a configuration $n = dv + Nn'$, we define 
\begin{align} \label{defS1}
\Psi_{\text{sym}}(\phi,a,dv+Nn') \propto \cdot \Psi(\phi,a-v,Nn')
\end{align}
The normalization factor must be $\sqrt{|{\{\beta|\partial\beta=0\}|}}{N^{-(|\text{faces}|-|\text{edges}|)/2}}$ to ensure unitarity.  This is eq. \ref{rho_inverse} in the main text.

\section{Mixed anomaly between $U_{\frac{1}{N}}(M^3)$ and $Z^{(1)}_e$} \label{app:mixed_anomaly_appendix}

One way to derive the mixed anomaly is to explicitly truncate $U_{\frac{1}{N}}(M^3)$ to 
an open $3$ dimensional subset $R\subset M^3$, and then conjugate the action of $Z^{(1)}_e$ by this truncated operator.  The result should look like $Z^{(1)}_e$ everywhere except near the $2$ dimensional boundary $\partial R$.  Physically, $R$ can be thought of as a finite $3$ dimensional region where we create the $1$-form SPT, and its $2$ dimensional boundary $\partial R$ is where the one-form symmetry forms an anomalous representation.  In the terminology of Else and Nayak \cite{Else_Nayak} (see also \cite{feng2025} for a general extension to higher-form symmetries), the restriction of the one-form symmetry $Z^{(1)}_e$ to $\partial R$ is an `effective boundary action' of the symmetry.  The symmetry operators must be supported on co-dimension one surfaces in $\partial R$, i.e. strings.  We expect the mixed anomaly between $U_{\frac{1}{N}}(M^3)$ and $Z^{(1)}_e$ to be encoded in this boundary action.  Specifically, we expect that the string operators implementing $Z^{(1)}_e$ in $\partial R$ do not commute, but rather follow the commutation relations of the string operators of the chiral $\Z_N$ anyon theory.

To explicitly see this, we will assume (just for the present argument) that we are on a cubic lattice $\Z^3$, with vertices having integer $x,y,z$ coordinates.  We will also take odd $N$, for simplicity.  The formalism of cup products and higher cup products extends to hypercubic lattices \cite{ChenTata2023}, so everything we have done so far can be ported over to this setting.  Let $R$ be the set of points $(x,y,z)$ with $z\leq 0$.  Then $\partial R$ is the set of points $(x,y,0)$, i.e. a $2$ dimensional square lattice $\Z^2$.  For concreteness, let us first consider the $1$-form action $Z_e^{(1)}(v)$ on $\Z^3$ with $v$ being $0$ everywhere except on edges connecting $(0,y,z)$ and $(1, y, z)$, for all $y,z$, where it is equal to $1$.  Then $Z_e^{(1)}(v)$ simply measures the fractional charge of Wilson lines extended along the $x$ direction.

Truncating $Z_e^{(1)}(v)$ to $R$, we obtain an operator $Z_e^{(1)}(v)_{\text{trunc}}$ which acts as follows.  Let $v'$ be $0$ everywhere except on edges connecting $(0,y,z)$ and $(1, y, z)$ for $z \leq 0$, where it is $1$.  Then
\begin{align*}
Z_e^{(1)}(v)_{\text{trunc}}:\Psi(\phi,a,n)\rightarrow \Psi(\phi, a-v', n)
\end{align*}
Note that there is no shift of $n\rightarrow n-dv'$, even though $dv' \neq 0$; indeed, there cannot be such a shift, as $n$ must remain valued in $N\Z$.  However, this is a valid truncation, in the sense that it acts the same way as $Z_e^{(1)}$ in the bulk of $R$, and as the identity away from $R$.

We now have
\begin{align} \label{trunc_conj}
U_{\frac{1}{N}}&(M^3) Z_e^{(1)}(v)_{\text{trunc}}U_{\frac{1}{N}}(M^3)^\dagger: \Psi(\phi,a,n) \rightarrow \nonumber \\ & \exp\left[\frac{2\pi i}{N}\int_{M^3} \left(dv'\cup\lr{a-d\phi} + \lr{a-d\phi}\cup dv' -v'\cup dv'\right)\right]\Psi(\phi,a-v',n) \nonumber \\
\propto \,& \exp\left[\frac{2\pi i}{N}\int_{M^3} \left(dv'\cup\lr{a-d\phi} + \lr{a-d\phi}\cup dv'\right)\right]\Psi(\phi,a-v',n)
\end{align}
We can express this operator as the composition of $Z^{(1)}_e(v)$ in the bulk of $R$ ($z<0$), together with an operator that acts non-trivially on the slab $0\leq z\leq 1$.  The non-trivial action involves the phase
\begin{align*}
\exp\left[\frac{2\pi i}{N}\int_{M^3} \left(dv'\cup\lr{a-d\phi} + \lr{a-d\phi}\cup dv'\right)\right]
\end{align*}
which, because $dv' \neq 0$ only on $2$-cells with $z$ coordinates between $0$ and $1$, is localized to the double layer slab $0 \leq z \leq 1$.

The form of eq. \ref{trunc_conj} is reminiscent of string operators in string net models.  Indeed, acting in the slab 
$0\leq z\leq 1$, this operator just shifts the value of $a$ by $1$ on edges $((0,y,0), (1,y,0))$, and multiplies by a phase proportional to $\lr{a-d\phi}$ on edges $((0,y,0),(0,y+1,0))$ and $((1,y,1), (1,y+1,1))$.  If we then define an analogous operator extended in the $x$ direction (i.e. exchange $x$ and $y$ in the definition of $v$), then these two operators will fail to commute precisely on edges of the form $((0,0,0),(1,0,0))$ and $((0,0,0),(0,1,0)$.  Furthermore, these two contributions are both equal to $\exp\left(\frac{2\pi i}{N}\right)$, so that the overall group commutator of these two operators is $\exp\left(\frac{4\pi i}{N}\right)$.  This is equal to the braiding of the generating anyon of the $\Z_N$ anyon theory with itself (recall that we are working with odd $N$ here, for simplicity).  One can also perform an exchange by using a T-junction process \cite{LevinWenEther}, which will involve an anti-commutation on only a single edge, leading to an exchange phase of $\exp\left(\frac{2\pi i}{N}\right)$.

Since the remaining anyons are just powers of the generating anyon, their braiding and exchange phases are reproduced as well.  We thus conclude that the anomalous one-form symmetry action on the two dimensional boundary of $R$ gives precisely the lines of the $\Z_N$ anyon theory, as expected.

\bibliographystyle{unsrt}
\bibliography{bibl}

\end{document}